\documentclass[12pt]{article}
\usepackage{epsf}
\usepackage{amsmath}
\usepackage{amsfonts}
\usepackage{amssymb}
\usepackage{graphicx}
\usepackage{color}
\usepackage{psfrag}
\usepackage{cite}

\usepackage{ifpdf}

\newcommand{\bmat}{\left(\begin{array}}
\newcommand{\emat}{\end{array}\right)}

\def\pim{{\rm Im\,}}

\def\yzero{\smash{\hbox{$y\kern-4pt\raise1pt\hbox{${}^\circ$}$}}}
\def\p{\partial}
\def\a{\alpha}
\def\b{\beta}

\def\d{\delta}
\def\be{\begin{equation}}
\def\ee{\end{equation}}
\def\bea{\begin{eqnarray}}
\def\eea{\end{eqnarray}}
\def\beq{\begin{equation}}
\def\eeq{\end{equation}}
\def\beqa{\begin{eqnarray}}
\def\eeqa{\end{eqnarray}}

\def\vt{\vartheta}

\def\-{\hphantom{-}}

\def\s2{\frac{1}{\sqrt2}}

\def\beq{\begin{equation}}
\def\eeq{\end{equation}}
\def\beqa{\begin{eqnarray}}
\def\eeqa{\end{eqnarray}}

\def\nn{\nonumber}

\def\IF{\relax{\rm I\kern-.18em F}}
\def\II{\relax{\rm I\kern-.18em I}}

\def\ca{{\cal A}}

\def\Dsl{\,\raise.15ex\hbox{/}\mkern-13.5mu D} 

\def\IC{{\bf C}}
\def\IS{{\bf S}}
\def\IR{{\bf R}}
\def\IZ{{\bf Z}}

\def\IT{{\bf T}}

\def\CM{{\cal M}}

\def\CO{{\cal O}}

\def\vt{\vartheta}

\def\eps{\epsilon}

\def\ca{{\cal A}}
\def\lam{\lambda}
\def\raw{\rightarrow}

\def\G{\Gamma}

\newcommand{\sbt}{\,\begin{picture}(-1,1)(-1,-3)\circle*{4}\end{picture}\ }

\catcode`\@=11   
\newdimen\@rotdimen
\newbox\@rotbox  

\def\@vspec#1{\special{ps:#1}}
\def\@rotstart#1{\@vspec{gsave currentpoint currentpoint translate
   #1 neg exch neg exch translate}}
\def\@rotfinish{\@vspec{currentpoint grestore moveto}}
\def\@rotr#1{\@rotdimen=\ht#1\advance\@rotdimen by\dp#1%
   \hbox to\@rotdimen{\hskip\ht#1\vbox to\wd#1{\@rotstart{90 rotate}%
   \box#1\vss}\hss}\@rotfinish}
\def\@rotl#1{\@rotdimen=\ht#1\advance\@rotdimen by\dp#1%
   \hbox to\@rotdimen{\vbox to\wd#1{\vskip\wd#1\@rotstart{270 rotate}%
   \box#1\vss}\hss}\@rotfinish}%
\def\@rotu#1{\@rotdimen=\ht#1\advance\@rotdimen by\dp#1%
   \hbox to\wd#1{\hskip\wd#1\vbox to\@rotdimen{\vskip\@rotdimen
   \@rotstart{-1 dup scale}\box#1\vss}\hss}\@rotfinish}%
\def\@rotf#1{\hbox to\wd#1{\hskip\wd#1\@rotstart{-1 1 scale}%
   \box#1\hss}\@rotfinish}%
\def\rotate{\@ifnextchar[{\@rotate}{\@rotate[l]}}
\def\@rotate[#1]#2{\setbox\@rotbox=\hbox{#2}\@nameuse{@rot#1}\@rotbox}

\catcode`\@=12

\topmargin
-1.5cm
\textwidth
15.5cm
\textheight
23.5cm
\oddsidemargin
0.7cm
\evensidemargin
1.2cm

\begin{document}

\makeatletter
\@addtoreset{equation}{section}
\makeatother
\renewcommand{\theequation}{\thesection.\arabic{equation}}
\pagestyle{empty}
\rightline{ IFT-UAM/CSIC-12-53}
\vspace{0.1cm}
\begin{center}
\LARGE{\bf Non-Abelian discrete gauge symmetries\\ in 4d string models
\\[12mm]}
\large{M. Berasaluce-Gonz\'alez$^{1,2}$, P. G. C\'amara$^{3}$, \\F. Marchesano$^{2}$, D. Regalado$^{1,2}$, A. M. Uranga$^2$\\[3mm]}
\footnotesize{${}^{1}$ Departamento de F\'{\i}sica Te\'orica,\\[-0.3em] 
Universidad Aut\'onoma de Madrid, 28049 Madrid\\
${}^2$ Instituto de F\'{\i}sica Te\'orica IFT-UAM/CSIC,\\[-0.3em] 
C/ Nicol\'as Cabrera 13-15, Universidad Aut\'onoma de Madrid, 28049 Madrid, Spain} \\
${}^3$ Departament de F\'{\i}sica Fonamental and Institut de Ci\`encies del Cosmos, Universitat de Barcelona, Mart\'{\i} i Franqu\`es, E-08028 Barcelona, Spain\\[2mm] 

\vspace*{3cm}

\small{\bf Abstract} \\[5mm]
\end{center}
\begin{center}
\begin{minipage}[h]{16.0cm}

We study the realization of non-Abelian discrete gauge symmetries in 4d field theory and string theory compactifications. The underlying structure generalizes the Abelian case, and follows from the interplay between gaugings of non-Abelian isometries of the scalar manifold and field identifications making axion-like fields periodic. We present several classes of string constructions realizing non-Abelian discrete gauge symmetries. In particular, compactifications with torsion homology classes, where non-Abelianity arises microscopically from the Hanany-Witten effect, or compactifications with non-Abelian discrete isometry groups, like twisted tori. We finally focus on the more interesting case of magnetized branes in toroidal compactifications and quotients thereof (and their heterotic and intersecting duals), in which the non-Abelian discrete gauge symmetries imply powerful selection rules for Yukawa couplings of charged matter fields. In particular, in MSSM-like models they correspond to discrete flavour symmetries constraining the quark and lepton mass matrices, as we show in specific examples.

\end{minipage}
\end{center}
\newpage
\setcounter{page}{1}
\pagestyle{plain}
\renewcommand{\thefootnote}{\arabic{footnote}}
\setcounter{footnote}{0}

\setcounter{tocdepth}{2}

\tableofcontents

\vspace*{1cm}


\section{Introduction}

Discrete symmetries are a key ingredient in particle physics, and especially in physics beyond the Standard Model. A prototypical example is the introduction of R-parity (or other similar Abelian symmetries) in the MSSM to forbid or suppress certain operators leading to exceedingly fast proton decay. Another fertile industry is the use of discrete (possibly non-Abelian) symmetries in flavour physics, to generate textures of quark and lepton masses and mixings.

Such discrete symmetries are thus introduced for phenomenological reasons, but their fundamental origin remains obscure. Of course they could be just accidental symmetries of the lowest-dimensional terms in the effective theory, but it is clearly important to consider them as possibly exact symmetries at the fundamental level. In this respect, there are diverse arguments strongly suggesting that global symmetries, either continuous or discrete, are violated by quantum gravitational effects, and hence cannot exist in any consistent quantum theory including gravity (see \cite{Banks:1988yz,Abbott:1989jw,Coleman:1989zu} for early viewpoints, and e.g.\cite{Kallosh:1995hi,Banks:2010zn} and references therein, for more recent discussions). This suggests that discrete symmetries should have a gauge nature in such theories \cite{Alford:1988sj,Krauss:1988zc,Alford:1989ch,Preskill:1990bm,Alford:1990mk,Alford:1990pt,
Alford:1991vr,Alford:1992yx}, in particular in string theory. 

The realization of discrete gauge symmetries in string theory is therefore an important topic. Abelian gauge symmetries, and their application to MSSM-like models have recently been explored in D-brane models in \cite{Camara:2011jg,BerasaluceGonzalez:2011wy,Ibanez:2012wg}, with nice agreement with the classification of anomaly-free discrete symmetries in  \cite{Ibanez:1991hv,Ibanez:1991pr} (see also \cite{Banks:1991xj,Ibanez:1992ji,Martin:1992mq,Dreiner:2005rd,Mohapatra:2007vd,Araki:2008ek,Lee:2010gv,Kappl:2010yu} and references therein; also \cite{Lebedev:2007hv} for attempts to implement R-parity in heterotic models).

In this paper we consider the realization of non-Abelian discrete gauge symmetries in field theory and string theory, and study the constraints that they impose on the Yukawa couplings of the theory. This requires generalizing the Abelian intuition that $\IZ_k$ symmetries can be constructed as remnants of U(1) gauge symmetries broken by vevs of charge $k$ fields. The non-Abelian generalization involves the gauging of non-Abelian isometries in the space of scalar fields of the theory, and its interplay with the field identifications (e.g. axion periodicities, or possibly more general dualities) in this scalar manifold. The analysis leads to a 4d Lagrangian formulation of (at least certain classes of) non-Abelian discrete gauge symmetries in terms of gauging of such `non-Abelian axions'. 

We also find several explicit realizations of this framework in the context of string theory, with particular focus on type II compactifications and orientifolds thereof (although much of the analysis holds more generally):

$\sbt$ First, extending the observation for 5d theories in \cite{Gukov:1998kn} (see also \cite{Burrington:2006uu,Burrington:2007mj}), we show that non-Abelian discrete gauge symmetries can arise from compactification of $p$-form fields along torsion homology classes with non-trivial relations (which we describe explicitly). We perform the dimensional reduction from 10d, and indeed find a 4d Lagrangian realizing non-Abelian discrete gauge symmetries, typically discrete Heisenberg groups, in terms of gaugings of non-Abelian axions. This generalizes the relation between torsion homology and discrete symmetries observed in the Abelian case in \cite{Camara:2011jg}.

$\sbt$ We also consider the realization of (possibly non-Abelian) discrete gauge symmetries from discrete isometries of the compactification space. Their general analysis is beyond our present scope, and we focus on the particular simple case of compactifications on twisted tori. Their realization in terms of gaugings allows to describe the discrete gauge symmetry in the language of gauging of non-Abelian axions. 

$\sbt$ Finally, we show that systems of magnetized branes on tori (or their heterotic or intersecting brane duals) in general enjoy non-Abelian discrete gauge symmetries, acting non-trivially on 4d charged matter fields. The resulting discrete symmetries have a Heisenberg-type structure and underlie some powerful selection rules for Yukawa couplings. These include those observed in \cite{Cremades:2003qj,Cremades:2004wa} (interpreted in terms of symmetries in \cite{Abe:2009vi}, see also \cite{Abe:2010iv}), and the rank one textures in certain MSSM-like models \cite{Cremades:2003qj}. Our analysis shows that these properties are not merely accidental but rather stem from genuine non-perturbatively exact discrete gauge symmetries present in the model.

The paper is organized as follows. In section \ref{sec:Abelian-gauging} we review Abelian discrete gauge symmetries, and emphasize their interpretation in terms of gaugings. The intuitions are subsequently generalized in section \ref{sec:nonAbelian} to realize non-Abelian discrete gauge symmetries in terms of gaugings of non-Abelian axions. In section \ref{sec:torsion-pforms} we realize non-Abelian discrete symmetries from compactification of $p$-form fields on geometries with torsion homology classes with relations. In section \ref{sec:isometries} we study non-Abelian discrete symmetries from compactification on geometries with discrete isometries, focusing on the illustrative case of twisted tori compactifications. In section \ref{sec:magnetized} we describe non-Abelian discrete symmetries in toroidal compactifications with magnetic fields. In section \ref{magnetized-toy} we derive the discrete symmetry in a toy situation of magnetization in a single $\IT^2$, and reproduce the constraints for Yukawa couplings appeared in  \cite{Cremades:2003qj,Cremades:2004wa}. In section \ref{sec:dim-red-t6} we derive the non-Abelian discrete symmetry from dimensional reduction of magnetized type I on $\IT^6$, and obtain the natural holomorphic variables in the effective action in section \ref{sec:holomorphic}. In section \ref{subsec:mssm} we describe such discrete symmetry for the MSSM-like model of  \cite{Cremades:2002qm,Cremades:2003qj}, and show it underlies the rank-one texture for Yukawa couplings in this model, which is thus exact even at the non-perturbative level. In section \ref{sec:instantons} we study non-perturbative instanton effects, and how they manage to preserve the non-Abelian discrete gauge symmetry. Section \ref{sec:conclusions} contains our final remarks, and several technical results are kept in appendices: appendix \ref{app:general} presents a generalization of the construction in section \ref{subsec:torsionforms}, appendix \ref{subsec:yukawas-kk} discusses the action of discrete symmetries of twisted tori on KK modes, and appendix \ref{app:details} details the derivation of eq. (\ref{relations-yukawas}).

The sections dealing with open strings in magnetized brane models are fairly self-contained. Hence, the reader interested just in the selection rules for charged matter fields in magnetized brane systems (or their duals), may get the relevant intuitions from section \ref{sec:Abelian-gauging} and jump onto section \ref{sec:magnetized}.

\section{Abelian discrete gauge symmetries and gaugings}
\label{sec:Abelian-gauging}

\subsection{Review of the Abelian case}
\label{sec:Abelian}

The basic action for a $\IZ_k$ discrete gauge symmetry (see \cite{Banks:2010zn} for a recent discussion)\footnote{See also \cite{Hellerman:2010fv} for a more formal viewpoint on theories with discrete gauge symmetries.} is 
\beqa
\int \, d^4x\, (\partial_\mu \phi \, -\, k A_\mu)^2\label{lagrangiano-Abeliano}
\eeqa
where the gauge field $A_1$ is normalized such that the minimum electric charge is 1, and $\phi$ is a scalar field (henceforth dubbed `axion') with a periodic identification
\beqa
\phi \simeq \phi+1
\label{discrete-equivalence}
\eeqa

The above Lagrangian can be dualized into an alternative $BF$ formulation, involving a 2-form and a (magnetic) gauge potential. Such formulation has been useful in the study of Abelian discrete gauge symmetries in string theory (e.g. \cite{Camara:2011jg,BerasaluceGonzalez:2011wy,Ibanez:2012wg}), but for our present purposes we however stick to the axion formulation. This form is largely inspired by considering $\phi$ to be the phase of a Higgs field with charge $k$ under a broken U(1) gauge group. However, we prefer to regard it just as a scalar, whose moduli space (locally given by $\IR$) has a continuous isometry
\beqa
\phi\to \phi+\epsilon
\eeqa
The action (\ref{lagrangiano-Abeliano}) describes the gauging of this isometry by a U(1), 
\beqa
A_\mu\, \to\,  A_\mu\, +\,\partial_\mu \lambda \quad , \quad \phi\, \to\, \phi\, +\, k\lambda
\label{Abelian-gauge-inv}
\eeqa
Before taking into account the periodicity (\ref{discrete-equivalence}), the value of $k$ could be removed by  rescaling $\phi$, and 
would not be relevant. The integer $k$ is thus properly interpreted as the winding number in the map between the $\IS^1$ of U(1) gauge transformations $e^{2\pi i\, \alpha}$ (with $\alpha\simeq \alpha+1$ due to charge quantization), and the $\IS^1$ parametrized by the axion $\phi$. The fact that $k$ is integer is a compatibility condition of the gauging by the U(1) with the pre-existing discrete equivalence (\ref{discrete-equivalence}). 

The gauging directly implements the field identification $\phi\simeq \phi+k$. On the other hand the discrete equivalence (\ref{discrete-equivalence}) corresponds to a `fractional' $1/k$ U(1) gauge transformation, namely a $\IZ_k$ gauge transformation. This perspective displays the close relation of the discrete gauge symmetry with the underlying field identification in the scalar manifold. More precisely, the discrete gauge symmetry is the group of field identifications in the scalar manifold modulo those already accounted for by the gauging. This intuition is the key to the non-Abelian generalization in the coming sections.

Theories with discrete gauge symmetries have sets of (possibly massive) charged particle states. These often provide a practical way to identify the discrete gauge symmetry in a given theory. In the case of the above $\IZ_k$ theory, charge $n$ particles with worldline $C$ are described as insertions of the line operator 
\begin{equation}
\mathcal{O}_{\rm particle}\sim e^{2\pi i n \int_C A_1} \label{particle}
\end{equation}
Their charge is conserved modulo $k$, since there are gauge invariant `instanton' vertices which create/annihilate sets of particles with total charge $k$, 
\beqa
e^{-2\pi i\, \phi} \, e^{2\pi i\, k\int_C A_1}\, =\, e^{-2\pi i\, \phi }\,  {\cal O}_{\rm particle(s)}
\label{instanton-operator}
\eeqa
describing an insertion $e^{-2\pi i\phi}$ at a point $P$, out of which a charge $k$ set of particles emerges along a worldline $C$ (i.e. $\partial C=P$). In many realizations, the above operators are induced in the 4d action by effects $e^{-S_{\rm inst}}$, non-perturbative in some suitable coupling, with $S_{\rm inst}=2\pi i\phi\, +\, \ldots$ linear in the gauged axion. The overall U(1) charge of $\mathcal{O}_{\rm particle(s)}$ is thus compensated by shifts of $S_{\rm inst.}$.

In addition, the theory contains $\IZ_k$ charged strings, described as the insertion of operators along a worldsheet $\Sigma$
\begin{equation}
\mathcal{O}_{\rm string}\sim e^{-2\pi i\, p\int_\Sigma B_2}\label{string}
\end{equation}
where $B_2$ is the 2-form dual to $\phi$, and $p$ is defined modulo $k$. A charge $n$ particle defined by (\ref{particle}) suffers a $\IZ_k$ discrete gauge transformation, $n\to n+p$, when moved around the charge $p$ string (\ref{string}), i.e. its wavefunction picks up an Aharonov-Bohm phase $e^{2\pi i\, pn/k}$. Conversely, a charge $p$ string looped around a charge $n$ particle picks up a phase $e^{2\pi i\, np/k}$. In more abstract terms, the amplitude associated to a charge $p$ string on a worldsheet $\Sigma$ and a charge $n$ particle on a worldline $C$ contains an Aharonov-Bohm phase
\beqa
\exp\left[\,2\pi i\, \frac{np}k L(\Sigma, C)\,\right]
\label{ab-phase}
\eeqa
where $L(\Sigma, C)$ is the so-called linking number of $\Sigma$ and $C$.

String charge is also conserved modulo $k$, since there are operators describing strings of total charge $k$ on worldsheets $\Sigma$ ending along a junction line $L$ ($\partial \Sigma=L$)
\beqa
e^{-2\pi i\, \int_LA_1} e^{2\pi i\, k\int_\Sigma B_2}
\eeqa
These ingredients have a natural yet more involved generalization to the non-Abelian case \cite{Alford:1989ch,Alford:1990mk,Alford:1990pt,
Alford:1991vr,Alford:1992yx} (see \cite{Lee:1994qg} for a review).

\subsection{The multiple Abelian case}
\label{sec:multiple-Abelian}

Before moving onto the non-Abelian case, let us sharpen our intuitions in a slightly more involved (yet Abelian) situation. Consider a theory with several U(1) gauge symmetries, labelled with an index $\alpha$, and several axions $\phi^a$, $a=1,\ldots, N$. The generalization of eq.~(\ref{lagrangiano-Abeliano}) is
\beqa
\mathcal{L}\supset\sum_{\alpha} \left( \,\partial_\mu \phi^a- k_\alpha{}^a A_\mu^\alpha\,\right)\, \left( \,\partial_\nu \phi^b- k_\alpha{}^b A_\nu^\alpha\,\right)\,\eta^{\mu\nu} \d_{ab}
\label{lag-mult-ab}
\eeqa
with integer $k_\alpha{}^a\in \IZ$. We take U(1) generators normalized such that charges are integer and axions have integer periodicity.

In general, it is not immediate to identify the surviving discrete gauge symmetry. In the literature this is usually done by `trial and error', by  
scanning through different integral linear combination of U(1) generators 
\beqa
Q\, =\, \sum_ \alpha c_\alpha Q_\alpha\quad \mbox{with $c_\alpha\in \IZ$ and g.c.d.($c_\alpha$)\ =\ 1}
\eeqa
and checking for the greatest common divisor of the U(1) axion couplings ${\textstyle\sum}_{\alpha} c_\alpha k_\alpha{}^a$.

There is however a systematic closed description of the surviving discrete gauge symmetry based on our earlier intuitions. For that aim, we consider the space spanned by the scalars $\phi^a$. This is a torus $\IT^N$ which we regard as $\IR^N/\Gamma$, with $\Gamma$ the lattice of translations defined by vectors of integer entries
\beqa
\Gamma\, =\, \{(r_1,\ldots,r_N)\, |\, r_a\in \IZ\,\}\label{lattice-scalar}
\eeqa
The Lagrangian (\ref{lag-mult-ab}) implies that U(1)$_\alpha$ gauge transformations act as translations in $\IR^N$ along the vectors $\vec{k}_\alpha$
\beqa
A_\alpha\to A_\alpha+d\lambda_\alpha\quad ;\quad \phi^a\to \phi^a+{\textstyle \sum}_\alpha k_\alpha{}^a\lambda_\alpha
\eeqa
For simplicity, we focus on the case where the number of axions and U(1) gauge symmetries is equal.\footnote{Generalization is straightforward. If the number $n$ of U(1)'s, is smaller than the number $N$ of scalars, we restrict to those scalars which actually shift: we consider the $\IR^n\subset\IR^N$ given by real linear combinations of the vectors $\vec{k}_\alpha$ (assumed linearly independent for simplicity), and the sublattice $\Gamma_n\subset\Gamma$ lying in this $\IR^n$, and proceed as above with $n$ playing the role of $N$.
For $\vec{k}_\alpha$ not linearly independent, we just eliminate the decoupled linear combinations of U(1)'s, and restart. Similarly if the number of U(1) gauge symmetries is larger than the number of scalars to start with.}
Finite U(1) gauge transformations leaving all charged fields invariant (i.e. gauge parameter $\lambda_\alpha=1$) act as discrete translations in $\IR^N$ by the integer vectors $\vec{k}_\alpha$, and therefore span a sublattice $\hat{\Gamma}\subset\Gamma$,
\beqa
\hat{\Gamma}\, =\, \langle \, \vec{k}_1,\ldots,\vec{k}_N\rangle_{\IZ}\,=\, \{\, {\textstyle \sum}_\alpha c_\alpha \vec{k}_\alpha\,|\, c_\alpha\in\IZ \,\}
\eeqa
Following our previous discussion for the single Abelian case, the discrete gauge symmetry is given by the set of identifications in the space of scalars modulo those implemented by the finite U(1) gauge symmetries, namely by the quotient
\beqa
{\bf P}=\frac{\Gamma}{\hat{\Gamma}}
\eeqa
As we will now see these intuitions generalize to the non-Abelian case as well.

\section{Non-Abelian discrete gauge symmetries and gaugings}
\label{sec:nonAbelian}

While the construction introduced above describes the well-known case of Abelian discrete gauge symmetries, it admits a natural generalization to the non-Abelian case. In the non-Abelian version instead of a single field we will have a whole set of scalars (dubbed `non-Abelian axions') which span a manifold with non-commuting isometries. This more general construction can also be regarded as a procedure to construct a Lagrangian formulation for (at least certain) non-Abelian discrete gauge theories.

\subsection{The scalar manifold}
\label{subsec:scalar-manifold}

Let ${\cal M}$ be the moduli space of $N$ scalars $\phi^a$, endowed with a metric $G_{ab}(\phi)$ with a set of (in general non-Abelian) continuous isometries with Killing vector fields $X_A=X_A^b\partial_b$. Under infinitesimal space-time independent isometry transformations the scalars transform as
\beqa
\phi^b\to \phi^b+\epsilon^A X_A^b\label{global}
\eeqa
and their kinetic term
\begin{equation}\label{kinetic}
\int d^4x\, G(\vec \phi)_{ab}\, \partial_\mu \phi^a\partial^\mu \phi^b  
\end{equation}
is invariant provided that $(\mathcal{L}_{X_A} G)_{ab}=0$. The Killing vector fields satisfy a Lie algebra
\beqa
[X_A,X_B]\, =\, f_{AB}{}^C X_C\label{algebra}
\eeqa
with $f_{AB}{}^{C}$ the structure constants and $[\ \ , \ ]$ the Lie Bracket.

Given the 4d Lagrangian (\ref{kinetic}) it is easy to guess how to implement a gauging analogous to eq.~(\ref{lag-mult-ab}), see eq.~(\ref{gauged}) below. Before doing that it is however useful to consider the scalar manifold ${\cal M}$ and try to understand which kind of metrics $G_{ab}(\phi)$ one may obtain in the case where all the fields $\phi^a$ are axions. This will allow in particular to rewrite (\ref{kinetic}) in a simpler form (namely eq.(\ref{genmetric}) below) which we will use extensively when reproducing non-Abelian discrete gauge symmetries from string theory setups.

In order to characterize the metric $G_{ab}$ it is useful to describe the manifold ${\cal M}$ in the language of group theory, as follows. Note that each Killing vector field describes a flow within $\CM$,  and so there is a natural action of the Lie group of isometries $\textrm{Iso}(\mathcal{M})$ on the scalar manifold $\CM$. We may then consider that $\textrm{Iso}(\mathcal{M})$ acts transitively on $\CM$,\footnote{If not, we may take the orbit $\CO_p$ created when $\textrm{Iso}(\mathcal{M})$ acts on a point $p \in \CM$, and then restrict the initial set of scalars $\phi^a$ to those that span $\CO_p$.} and so identify $\CM$ with the coset $K_p\backslash\textrm{Iso}(\mathcal{M})$, with $K_p$ the stabilizer or little group of an arbitrary point $p \in \CM$. Therefore we may apply the usual procedure (see for instance appendix A.4 of \cite{648696}) for building a Riemannian metric $G_{ab}(\phi)$ for $\CM$ in terms of the elements of $\textrm{Iso}(\mathcal{M})$ and $K_p$.

In general, the quotient $K_p\backslash\textrm{Iso}(\mathcal{M})$ will not be a Lie group itself: for this it is necessary that $K_p$ is a normal subgroup of $\textrm{Iso}(\mathcal{M})$. However, if ${\cal M}$ is parametrizes the vevs of {\em only} axion-like scalars, the choice of ${\cal M}$ as a Lie group is quite natural. Indeed, for an `axionic manifold' ${\cal M}$ the number of independent shift symmetries at any point should equal the dimension of $\CM$. This is automatically satisfied if ${\cal M}$ is a Lie group, since in this case we can identify each axion with an element of the Lie algebra of the group $\CM$, while the continuous shift symmetry corresponds to the one-parameter subgroup generated by such Lie algebra element. Hence, in the following we will consider the case where our axionic manifold $\CM$ is a Lie group.\footnote{In general we would expect that a coset $\CM$ that is not a Lie group but is nevertheless a parallelizable manifold could also qualify as an axionic manifold. We are nevertheless unaware of any example of this kind arising from a string compactification, and so this possibility will not be analyzed here.}

In the case that $\CM$ is a Lie group we can systematically build an affine representation of $\mathcal{M}$ acting on the plane $\IR^{N+1}$, with $N = {\rm dim\, } \CM$. For this construction, familiar from the description of twisted tori geometries, we first consider the affine plane $\IR^{N+1}$ described by vectors
\begin{equation}
\vec v=\begin{pmatrix}\vec \phi\\ 1\end{pmatrix}\label{affine}
\end{equation}
as well as a vector $\vec\epsilon\in\IR^{N}$ that parametrizes an element of the Lie algebra of $\CM$. Second, we consider the adjoint representation of $\textrm{Lie}(\mathcal{M})$, given by $(\textrm{ad}_{\vec \epsilon})_b{}^c=\epsilon^a f_{ab}{}^c$, and construct the matrices
\begin{equation}
\mathfrak{g}(\vec\epsilon)=\begin{pmatrix}\frac12\textrm{ad}_{\vec \epsilon}& \vec\epsilon\\ 0& 0\end{pmatrix}
\end{equation}
which provide a faithful $(N+1)$-dimensional linear representation of $\textrm{Lie}(\mathcal{M})\subset\mathfrak{iso}(\mathcal{M})$. Taking the exponential map, we obtain
\begin{equation}
g(\vec\epsilon)=\begin{pmatrix}e^{\frac12\textrm{ad}_{\vec \epsilon}}& 2\, \textrm{ad}^{-1}_{\vec \epsilon}(e^{\frac12\textrm{ad}_{\vec \epsilon}}-\mathbb{I}_{n\times n})\, \vec\epsilon\\ 0& 1\end{pmatrix}\label{gaff}
\end{equation}
where $\epsilon^a$ now parametrize arbitrarily large translations in $\mathcal{M}$. Finally, we can build an explicit expression for the metric $G_{ab}(\phi)$ in terms of the right-invariant 1-forms $\eta^a$, which are defined as
\beqa
(dg \cdot g^{-1})(\vec\phi)=\eta^a(\vec\phi)\, t_a
\label{eta-first}
\eeqa
with $t_a$ the generators of $\textrm{Lie}(\mathcal{M})$. We then obtain that the metric for $\CM$ is such that
\begin{equation}
\int d^4x\, G_{ab}(\vec \phi)\,\partial^\mu\phi^a\partial_\mu\phi^b=\int d^4x\, \mathcal{P}_{ab} \eta^a \cdot \eta^b
\label{genmetric}
\end{equation}
where $\mathcal{P}_{ab}$ is the metric in the tangent space of $\mathcal{M}$, and so independent of $\phi$, while $\eta^a \cdot \eta^b \equiv \eta^{\mu\nu} \eta^a_\mu \eta^b_\nu$ with $\eta^{\mu\nu}$ the 4d Minkowski metric. Notice that this expression is automatically invariant under continuous right-translations by group elements  $g(\vec\phi)\, \to\, g(\vec\phi)g(\vec\epsilon)$, and so it indeed respects the axionic shift symmetries. 

A particularly relevant case to forthcoming applications is when $\textrm{Lie}(\mathcal{M})$ is a 2-step nilpotent algebra (see \cite{Andriot:2010ju} for a recent review). In this case we have that $e^{\frac12\textrm{ad}_{\vec \epsilon}}=1+\frac12\textrm{ad}_{\vec \epsilon}$ and so eq.~(\ref{gaff}) reduces to
\begin{equation}
g(\vec \epsilon)=\begin{pmatrix}1+\frac12\textrm{ad}_{\vec \epsilon}& \vec\epsilon\\ 0& 1\end{pmatrix}\label{gnil}
\end{equation}
Then, applying eq.~(\ref{eta-first}) we obtain
\begin{equation}
\eta^a_\mu=\partial_\mu\phi^a+\frac12 f_{bc}{}^a\phi^b\, \partial_\mu\phi^c
\label{nilmetric}
\end{equation}
yielding a particularly simple expression for the right-invariant forms $\eta^a$ and hence for the metric in (\ref{genmetric}). 

Since the above construction is general it is important to note that, unless ${\rm Lie}(\CM)$ is semi-simple, ${\cal M}$ will be a non-compact manifold which is unsuitable to describe the moduli space of axionic-like scalars. We may however make this moduli space compact by taking its quotient by a lattice $\Gamma \subset \CM$. This is in fact something quite common in string theory, where moduli spaces are quotients of the form ${\tilde {\cal M}}= {\cal M}/\Gamma$, with $\Gamma$ a discrete subgroup of $\textrm{Iso}(\mathcal{M})$ that takes into account the dualities of the theory. A well-known example is the 10d axio-dilaton coupling $\tau$ of type IIB theory, whose moduli space is not $\CM_\tau = SO(2)\backslash SL(2,\IR)$ but rather ${\tilde {\cal M}}_\tau=SO(2)\backslash SL(2,\IR)/SL(2,\IZ)$ once S-duality has been taken into account.\footnote{As a slightly more involved example, we may reconsider the multiple Abelian case in subsection \ref{sec:multiple-Abelian}. Before taking the quotient by the lattice (\ref{lattice-scalar}), the scalar manifold is $\mathcal{M} = \IR^{N}$ and its isometry group is the Euclidean group, $\textrm{Iso}(\mathcal{M}) = \IR^{N} \rtimes O(N)$. Since the action of $\textrm{Iso}(\mathcal{M})$ on $\CM$ is transitive and the little group of any point of $\CM$ is $O(N)$, $\mathcal{M}$ can be identified with the quotient $O(N)\backslash\textrm{Iso}(\mathcal{M})$, which is nothing but the group of translations in $\mathcal{M}$. Finally, this space is made compact by taking the quotient $\tilde{\mathcal{M}} =\mathcal{M}/\Gamma$, with $\Gamma$ a group of discrete translations. }

Going back to the general case, if $\Gamma$ is cocompact, namely if there is a subset $X\subset \CM$ such that the image of $X$ under the action of $\Gamma$ covers the entire  $\CM$, then $\tilde{\mathcal{M}}= \mathcal{M}/\Gamma$ is compact. Finding such a lattice is in general a complicate task and its existence is not guaranteed. However, if $\mathcal{M}$ is a nilpotent Lie group it is enough to require that the structure constants are integer in some particular basis and that they satisfy $f_{ab}{}^a=0$ \cite{Malcev}. For the time being we will assume that such cocompact $\Gamma$ exists, but ignore its effect until subsection \ref{subsec:heisenberg-example}.

\subsection{The gauging}
\label{subsec:gauging}

Let us now write a 4d Lagrangian describing a set of non-commuting U(1) gauge symmetries that gauge some of the isometries of ${\cal M}$, ignoring the effect of the discrete lattice $\Gamma$. To describe such gauging, instead of (\ref{global}), we consider infinitesimal space-time dependent isometry transformations
\beqa
\phi^b\to \phi^b+\epsilon^A(x) X_A^b\label{infini}
\eeqa
where $x$ represents the set of 4d coordinates. Invariance of the action under local transformations becomes manifest once we introduce the corresponding set of gauge fields (see e.g. \cite{Hull:1990ms}). We have the generalization of (\ref{lag-mult-ab})
\beqa
\int d^4x\, G_{ab}(\phi) \left(\, \partial_\mu\phi^a\, -\, k_\alpha{}^a A_\mu^\alpha\, \right)\left(\, \partial_\nu\phi^b\, -\, k_\beta{}^b A_\nu^\beta\, \right)\eta^{\mu\nu}\label{gauged}
\eeqa
where the set of vector fields $\{k_\alpha\}$ is similar to the above $\{X_A\}$, but not necessarily identical due to relative normalizations to be discussed in the next subsections. In order for this action to be invariant under the infinitesimal isometry transformations (\ref{infini}), covariant derivatives have to transform as
\begin{equation}
\partial_\mu\phi^a-k_\alpha{}^aA^\alpha_\mu\ \to \ (\delta^a_b+\epsilon^A\partial_bX_A^a)(\partial_\mu\phi^b-k_\alpha{}^bA_\mu^\alpha)
\end{equation}
which means that the gauge fields $A_\mu^\alpha$ transform as
\begin{equation}
k_\alpha{}^a\, A_\mu^\alpha\ \to \ k_\alpha{}^a\, A_\mu^\alpha\, + \, X_C^a\partial_\mu\epsilon^C\, +\, f^C{}_{AB}\, X_C^a\, (X^{-1})_b^A\, A_\mu^\beta\, k_\beta{}^b\, \epsilon^B\label{gauge-transformation}
\end{equation}

As in the previous section, let us focus on the case where $\mathcal{M}$ is a Lie group. For notational simplicity, we will assume that all the right isometries of $\mathcal{M}$ are gauged. It is easy to see that the right-invariant 1-forms are now given by
\beqa
(D g \cdot g^{-1})(\vec\phi)=\eta^a(\vec\phi)\, t_a\label{eta}\quad {\rm with} \;\, D g=d g-t_a k^a_\alpha A_\mu^\alpha,
\eeqa
and so are built by performing the replacement $dg \raw Dg$ everywhere. In terms of these new 1-forms the action is still given by 
\begin{equation}
\int d^4x\, \mathcal{P}_{ab}\eta^a\cdot \eta^b\label{metrica-gauged}
\end{equation}
As before, for the particular case of 2-step nilpotent groups things simplify and these right-invariant 1-forms read
\begin{equation}
\eta^a_\mu=\partial_\mu\phi^a-k_\alpha{}^a A^\alpha_\mu+\frac12 f_{bc}{}^a\phi^b\, (\partial_\mu\phi^c-k_\beta{}^c A^\beta_\mu)\label{1forms-gauged}
\end{equation}
and so under a space-time dependent right-translation $g(\vec\phi)\to g(\vec \phi)g(\vec\epsilon)$ gauge fields transform as
\begin{equation}
k_\alpha{}^aA^\alpha_\mu t_a\ \to \ k_\alpha{}^aA^\alpha_\mu t_a g(\vec\epsilon) + g(\vec\phi)\partial_\mu g(\vec\epsilon)
\end{equation}

\subsection{A simple example}
\label{subsec:heisenberg-example}

The above construction provides the Lagrangian for a massive non-Abelian gauge symmetry, but it still does not reveal potential residual discrete gauge symmetry. To proceed further and make the discussion concrete, we introduce here an example of scalar manifold $\mathcal{M}$ and lattice $\Gamma$ whose gauging leads to a non-Abelian discrete symmetry group. The example is constructed using the Heisenberg group, $\mathcal{M}=\mathcal{H}_3(\IR)$, and will be realized in several physical systems in coming sections. In the next subsection we then extend the discussion to the general case.

Thus, we consider the 3-dimensional Heisenberg group as generated by matrices of the form
\begin{equation}
g(\vec\epsilon)=\begin{pmatrix}
1&0&0&\epsilon^1\\
0&1&0&\epsilon^2\\
-\frac{M}{2}\epsilon^2 & \frac{M}{2}\epsilon^1 & 1 & \epsilon^3\\
0 & 0 & 0 & 1\end{pmatrix}\label{group}
\end{equation}
with $M$ an integer. The associated Lie algebra is
\begin{equation}
[t_1,t_2]=M t_3\label{alg}
\end{equation}
where $t_1$, $t_2$ and $t_3$ are the elements of the algebra that generate the 1-dimensional subgroups parametrized by $\epsilon^1$, $\epsilon^2$ and $\epsilon^3$. The right-invariant 1-forms are given by eq.~(\ref{nilmetric}), which in this particular case corresponds to
\begin{equation}
\eta^1_\mu=\partial_\mu \phi^1\ , \qquad \eta^2_\mu=\partial_\mu \phi^2\ , \qquad \eta^3_\mu=\partial_\mu \phi^3 + \frac{M}{2}(\phi^1\partial_\mu \phi^2 - \phi^2\partial_\mu \phi^1)\label{eta-heis-ung}
\end{equation}
in terms of which the metric of $\mathcal{M}$ is given by the r.h.s. of (\ref{genmetric}). 

Since $\mathcal{M}$ is non-compact, we take our axionic moduli space to be given by the compact coset $\tilde{\mathcal{M}}= \mathcal{H}_3(\IR)/\Gamma$ where, for concreteness, we take the cocompact lattice $\Gamma\subset \mathcal{H}_3(\IR)$ to be generated by $(\phi^1,\phi^2,\phi^3)=(n_1,n_2,n_3)$ with $n_i\in\IZ$, namely by the discrete transformations
\begin{align}
\Gamma(1,0,0) \ : \qquad \phi^1&\to\phi^1+1 \ , \qquad \phi^3 \to \phi^3 - \frac{M}{2}\phi^2\ ,  \label{discrete-example}\\
\Gamma(0,1,0) \ : \qquad \phi^2&\to\phi^2+1 \ , \qquad \phi^3 \to \phi^3 + \frac{M}{2}\phi^1\ , \nonumber\\
\Gamma(0,0,1) \ : \qquad \phi^3&\to\phi^3+1\ . \nonumber \end{align}

We can gauge the right isometries of $\tilde{\mathcal{M}}$ following the general procedure described in the previous subsection. Thus, we introduce a set of U(1) gauge bosons $A^\alpha_\mu$, $\alpha=1,2,3$, and replace the right-invariant 1-forms (\ref{eta-heis-ung}) by their gauged counterparts eq.~(\ref{1forms-gauged}), which in this particular case read 
\begin{align}
&\eta^1_\mu=\partial_\mu \phi^1-k_1A^1_\mu\ , \qquad\qquad \eta^2_\mu=\partial_\mu \phi^2-k_2A^2_\mu\ , \label{gauged-forms-example}\\
&\eta^3_\mu=\partial_\mu \phi^3-k_3A^3_\mu +\frac{M}{2}\left[\phi^1(\partial_\mu \phi^2-k_2A^2_\mu) - \phi^2(\partial_\mu \phi^1-k_1A^1_\mu)\right]\nonumber
\end{align}
with $k_\alpha\in \IZ$, $\alpha=1,2,3$. 

After the gauging, U(1) gauge transformations of the gauge bosons $A^\alpha_\mu$ induce non-trivial shifts on the scalars
\begin{align}
& A^1_\mu \to A^1_\mu + \partial_\mu\lambda^1 \ ,  \qquad \qquad A^2_\mu \to A^2_\mu + \partial_\mu\lambda^2\ ,  \label{discrete-gauge-example}\\
&A^3_\mu \to A^3_\mu + \partial_\mu\lambda^3 +\frac{M k_1k_2}{2k_3}\left(\lambda^2A^1_\mu-\lambda^1A^2_\mu\right)+\frac{M}{2k_3}(k_2\phi^1\partial_\mu \lambda^2-k_1\phi^2\partial_\mu\lambda^1)\ ,\nn \\
&\phi^1\to\phi^1+k_1\lambda^1 \ , \qquad \phi^2\to\phi^2+k_2\lambda^2 \ ,\qquad
\phi^3 \to \phi^3 + \frac{M}{2}(k_2\phi^1\lambda^2-k_1\phi^2\lambda^1)+k_3\lambda^3\ .\nn \end{align}
Compatibility of these transformations with (\ref{discrete-example}) then leads to a set of non-commuting $\IZ_{k_\alpha}$ discrete gauge symmetries. Indeed, the gauge symmetry is given by the set of identifications (\ref{discrete-example}) modulo these finite gauge transformations, in analogy with the Abelian case. For instance, for $k_1=k_2=k_3=k\in \IZ$ and $M=1$ we have that the discrete gauge symmetry is given by ${\bf P}=(\IZ_k\times\IZ_k)\rtimes \IZ_k$, with generators  $\tilde T_1$, $\tilde T_2$ and $\tilde T_3$ satisfying
\begin{equation}
\tilde T_1^k=\tilde T_2^k=\tilde T_3^k=1 \ , \qquad \tilde T_1\tilde T_2=\tilde T_3\tilde T_2\tilde T_1\label{finite-heisenberg}
\end{equation}
For $k=2$ this is isomorphic to the dihedral group, ${\bf P}\simeq\textrm{Dih}_4$, whereas for $k=3$ the discrete symmetry group is ${\bf P}\simeq\Delta(27)$.

\subsection{The discrete gauge symmetry}
\label{subsec:discrete-symmetry}

To obtain the non-Abelian discrete gauge symmetry group in the above example we have closely followed a similar reasoning to the one that we used for Abelian discrete gauge symmetries. Indeed, we have seen that gauge transformations span a lattice $\hat\Gamma\subset \tilde{\mathcal{M}}$ and in order to gauge the left isometries of $\tilde{\mathcal{M}}$ it is enough to specify such a lattice. As in the Abelian case, the discrete gauge symmetry arises when we take into account the group $\Gamma$ of scalar field identifications; namely when we specify the periodicities of the isometries generated by $X_A$ and compare them with those of the gauge transformations (\ref{gauge-transformation}), generated by $k_\alpha$. Thus, once $\Gamma$ is taken into account, a non-trivial compatibility condition for the gauging arises. 

The discrete gauge symmetry of the theory is
\beqa
{\bf P}=\frac{\Gamma}{\hat\Gamma}
\eeqa
Fields charged under the original U(1) symmetries end up in some representation of this discrete gauge symmetry (whether they are massless fields or not), as we later on discuss in explicit string theory examples. 

\section{Torsion $p$-forms and discrete gauge symmetries}
\label{sec:torsion-pforms}

Gauged shift symmetries, and thus discrete gauge symmetries, are ubiquitous in 4d string theory models. This is particularly manifest in D-brane models, where Abelian discrete gauge symmetries arise from the coupling of D-brane U(1)'s to axion fields \cite{BerasaluceGonzalez:2011wy}. 
As pointed out in \cite{Camara:2011jg}, another source of discrete gauge symmetries appears by considering compactification manifolds with torsion in homology. As we will discuss in this section these two frameworks are directly related, and the latter can be easily generalized to describe non-Abelian discrete symmetries. In fact, we will show that dimensional reduction of type IIB supergravity on a manifold with torsion produces the 4d Lagrangian of
non-Abelian discrete gauge symmetries constructed in the previous section. 

\subsection{Abelian discrete gauge symmetries and torsion homology}
\label{sec:Abelian-torsion}

Before describing the non-Abelian case let us review the relation of Abelian discrete gauge symmetries to torsion classes \cite{Camara:2011jg}. As mentioned in section \ref{sec:Abelian}, a practical way to identify discrete gauge symmetries is to tag a set of $\IZ_k$ charged particles and $\IZ_k$ charged strings inducing relative holonomies on each other via the Aharonov-Bohm phase (\ref{ab-phase}). In string theory compactifications, we thus search for dynamical objects in the higher-dimensional theory that lead to Aharonov-Bohm strings and particles in the 4d effective theory. 

\subsubsection{Aharonov-Bohm strings and particles from torsion}

A simple way to obtain Aharonov-Bohm strings and particles in type II vacua is to consider D-branes or NS-branes wrapped on $p$-cycles of the compactification manifold, the inequivalent possibilities being classified in terms of homology. In general, the homology group of a $D$-dimensional manifold $X_D$ consists of a free part, given by $b_p\equiv \textrm{dim }H_r(X_D,\IR)$ copies of $\IZ$, and a torsion part, given by a set of finite $\IZ_k$ groups,
\begin{equation}
H_p(X_D,\IZ)\, =\, H_p^{\rm free}(X_D,\IZ)\, \oplus\, \textrm{Tor }H_p(X_D,\IZ)\, =\, \IZ^{b_p} \oplus\, (\IZ_{k_1}\oplus\ldots\oplus\IZ_{k_n})
\end{equation}
It has been argued in \cite{Camara:2011jg} that 4d Aharonov-Bohm strings and particles arising from a compactification in $X_D$ are associated to the torsion part of the corresponding homology lattice. This is based on the observation that if we wrap a $p$-brane on a torsion $p$-cycle $\pi_p^{\rm tor}$ and a dual $(D-p)$-brane on a torsion $(D-p-1)$-cycle $\pi^{\rm tor}_{D-p-1}$ then we will have a 4d particle and string, respectively, that induce fractional holonomies on each other proportional to the torsion linking number $L([\pi_p^{\rm tor}],[\pi^{\rm tor}_{D-p-1}])$ in the internal dimensions. Such torsion linking number is one of the main topological invariants that can be defined for the torsion homology classes of $X_D$, and it univocally relates torsion classes of $p$-cycles to torsion classes of $(D-p-1)$-cycles, such that 
\begin{equation}
\textrm{Tor }H_p(X_D,\IZ)\, \simeq\, \textrm{Tor }H_{D-p-1}(X_D,\IZ)\label{torsion-classes}
\end{equation}

Let us be more specific and consider M-theory compactified on a manifold $X_7$ with $G_2$ holonomy. Gauge symmetries in the 4d effective theory arise from the M-theory 3-form $A_3$ and are classified by elements of $H_2(X_7,\IZ)$. On the one hand, elements belonging to the free part of $H_2(X_7,\IZ)$ are in one-to-one correspondence with harmonic 2-forms in $X_7$ so, upon expanding $A_3$ in such 2-forms, we obtain standard U(1) gauge symmetries in the 4d effective theory. On the other hand, elements that belong to $\textrm{Tor\,}H_2(X_7,\IZ)$ must correspond to discrete $\IZ_{k_i}$ gauge symmetries.\footnote{If the manifold has discrete isometries, there can be in addition discrete gauge symmetries coming from the metric. We touch upon them in section \ref{sec:isometries}.} This can be seen from the fact that M2-branes wrapping torsion 2-cycles lead to Aharanov-Bohm particles in 4d, whereas M5-branes wrapping the dual torsion 4-cycles (which exist because of eq.(\ref{torsion-classes})) lead to 4d Aharanov-Bohm strings. 

Indeed, let us consider an M2-brane wrapping a $\IZ_k$ torsion 2-cycle $\pi_2^{\rm tor}$ and with 4d worldline $C$, as well as a 4d  string with worldsheet $\Sigma$ that arises from an M5-brane wrapping a $\IZ_k$ torsion 4-cycle $\pi_4^{\rm tor}$ of $X_7$. Following  \cite{Camara:2011jg}, one can see that the holonomy that these two objects induce on each other is given by
\bea
\label{hol1}
\frac{1}{2\pi i}\textrm{log }\left[\textrm{hol}(\Sigma,C)\right]& \stackrel{{\rm mod\, } 1}{=}& \int_{C \times \pi_2^{\rm tor}} A_3\, =\, \frac{1}{k} \int_{D \times k\pi_2^{\rm tor}} F_4\, =\, \frac{1}{k} \int_{D \times S_3} \delta_5\\
& \stackrel{{\rm mod\, } 1}{=} & \int_{\Sigma \times \pi_4^{\rm tor}} A_6\, =\, \frac{1}{k} \int_{B \times k\pi_4^{\rm tor}} F_7\, =\, \frac{1}{k} \int_{B \times S_5} \delta_8
\label{hol2}
\eea
The upper chain of equalities represent the Aharanov-Bohm effect that a 4d string creates on a 4d particle circling around it with a path $C = \p D$. Indeed, the M5-brane that becomes a 4d string will create a flux $F_4$ via backreaction, and we should integrate the corresponding potential $A_3$ on the M2-brane worldvolume $C \times  \pi_2^{\rm tor}$ to compute the induced holonomy on the 4d particle. The computation is then carried by applying Stokes' theorem and by noticing that because $\pi_2^{\rm tor}$ is $k$-torsion there is a 3-chain $S_3$ such that $\p S_3 = k \pi_2^{\rm tor}$, and that $dF_4 = \d_5$ with $\d_5$ a bump 5-form transverse to the M5-brane worldvolume $\Sigma \times \pi_4^{\rm tor}$. Similarly, the lower chain represents the holonomy created by the 4d particle on a 4d string surrounding it with $\Sigma = \p B$, with now $\p S_5 = k  \pi_4^{\rm tor}$ and $dF_7 = \d_8$. Notice that the integral of a bump function like $\d_5$ or $\d_8$ is always an integer, and so we end up with a fractional holonomy of the form exp ($2\pi i \ell/k)$ with $\ell \in \IZ$. One can see that the integer $\ell$ in eqs.~(\ref{hol1}) and (\ref{hol2}) is the same integer mod $k$, since both quantities in the r.h.s. are the definition of the torsion linking number $L([\pi_2^{\rm tor}],[\pi^{\rm tor}_4])$ multiplied by the 4d linking number $L(\Sigma,C)$ of eq.(\ref{ab-phase}). 

To summarize, one finds that the Aharanov-Bohm phase that an M2-brane and an M5-brane wrapped on torsion cycles create on each other is given by
\be
\exp\left[\,2\pi i\, L([\pi_2^{\rm tor}],[\pi^{\rm tor}_4])\cdot L(\Sigma, C)\right]
\ee 
Comparing with eq.(\ref{ab-phase}), we can identify $L([\pi_2^{\rm tor}],[\pi^{\rm tor}_4]) = np/k$, and so the charges $n$ and $p$ of the 4d objects correspond in the higher dimensional M-theory picture to choose torsion cycles with appropriate linking numbers. 

This M-theory picture allows to reinterpret the Abelian discrete gauge symmetries that arise in type IIA  compactifications with intersecting D6-branes \cite{BerasaluceGonzalez:2011wy}. Indeed, if the $G_2$ manifold $X_7$ admits a weakly coupled type IIA limit with D6-branes, some of the U(1) symmetries classified by $H_2 (X_7,\IZ)$ are downlifted to U(1) symmetries localized at D6-branes. Massless 4d particles charged under such U(1)'s, which in type IIA are open strings at the D6-brane intersections, correspond to M2-branes wrapping collapsed 2-cycles of $X_7$. The U(1) gauge symmetries that in M-theory are related to $H_2^{\rm free}(X_7,\IZ)$ become in type IIA D6-brane U(1) symmetries without any axion coupling, while those discrete gauge symmetries related to $\textrm{Tor\,}H_2(X_7,\IZ)$ become D6-brane U(1)'s broken to $\IZ_k$ through axion couplings. Consequently, massless 4d particles are charged under the unbroken U(1)'s if they are M2-branes wrapped on non-torsional 2-cycles, while particles that only have $\IZ_k$ charges correspond to M2-branes wrapping collapsed torsional 2-cycles of $X_7$.

This M-theory perspective provides also a geometrization of the instanton contribution structure (\ref{instanton-operator}), as follows. Consider a set of particles $\psi_i$ with $\IZ_k$ charges, namely a set of M2-branes wrapping torsion 2-cycles $D_i$; whenever the total homology charge of the combination is zero (in particular, the torsion classes add up to a trivial class) there exists a 3-chain $S$ connecting them ($\partial S=\sum_i D_i$). An M2-brane wrapped on $S$ would describe an instanton effect on the 4d theory, but it contains open holes. A completely consistent instanton can be obtained by glueing M2-branes on $D_i$, emerging from the instanton from the 4d perspective. This is precisely the dressed instanton structure (\ref{instanton-operator}) with ${\cal O}=\prod_i \psi_i$. Also, this is the M-theory picture of a D2-brane instanton with insertions of 4d charged matter multiplets, observed in \cite{Blumenhagen:2006xt,Ibanez:2006da,Florea:2006si}.

\subsubsection{Torsion and dimensional reduction}

Interestingly, this geometrical picture that relates torsion to discrete gauge symmetries can also be made manifest by means of dimensional reduction \cite{Camara:2011jg}. For this, we need to associate to each generator of $\textrm{Tor\,}H_p(X_D,\IZ)$ a differential $p$-form which is also an eigenform of the Laplacian, just like we do when we associate harmonic $p$-forms to the generators of $H_p^{\rm free}(X_D,\IZ)$. In the case of torsion groups, however, these eigenforms must have a non-zero eigenvalue and in order to reproduce the topological information of $\textrm{Tor\,}H_p(X_D,\IZ)$ we must consider non-closed $p$-forms satisfying specific relations. More precisely, given the generators of $\textrm{Tor }H_p(X_D,\IZ)$ and $\textrm{Tor }H_{D-p-1}(X_D,\IZ)$ we consider non-closed $p$- and $(D-p-1)$-forms $\omega_\a$ and $\a^\beta$ such that  \cite{Camara:2011jg}
\begin{equation}
d\omega_\alpha=k_\alpha{}^\beta\beta_\beta\qquad\qquad d\alpha^\beta=(-1)^{D-p}k^\beta{}_\alpha\tilde{\omega}^\alpha 
\label{general-relations}
\end{equation}
where $\b_\b$ and $\tilde{\omega}^\alpha$ are exact eigenforms of the Laplacian which are trivial in de Rham cohomology but represent non-trivial elements of $H^{p+1}(X_D,\IZ)$ and $H^{D-p}(X_D,\IZ)$, respectively. Moreover, $k_\alpha{}^\beta\in \IZ$ must be given by
\begin{equation}
L([\pi^{\rm tor}_{p,\alpha}],[\pi_{D-p-1}^{{\rm tor},\beta}])=(k^{-1})_\alpha{}^\beta
\end{equation}
so that it contains the topological information of the torsion cycles that these eigenforms are related to. Finally, the integral of these forms satisfy
\begin{equation}
\int_{X_D}\alpha^\rho\wedge\beta_\sigma = \int_{X_D}   \tilde{\omega}^\rho \wedge \omega_\sigma =
\delta^\rho_\sigma
\end{equation}

Including this set of non-harmonic eigenforms when performing the dimensional reduction allows to reproduce the 4d Lagrangian for Abelian discrete gauge symmetries, and in particular displays the gauging structure discussed in section \ref{sec:Abelian}. Indeed, taking again the above example of M-theory on 7-manifolds, for each torsion 2-cycle we need to consider an exact 3-form $\alpha_3$ and a non-closed 2-form $\omega_2$, with $d\omega_2=k\beta_3$ and $k\in\IZ$. Expanding the M-theory 3-form $A_3$ in such eigenforms we obtain
\beqa
A_3\, =\, \phi (x^\mu)\wedge\beta_3 \, +\, A_1(x^\mu)\wedge \omega_2\, + \, \ldots
\eeqa
namely, a 4d U(1) gauge boson $A_1$ and a 4d scalar $\phi$. One can check that the gauge transformation (\ref{Abelian-gauge-inv}) shifts $A_3$ by the exact 3-form $d(\lam \omega_2)$ and so it indeed leaves any 4d quantity invariant. In particular we have that
\begin{equation}
dA_3\, =\, \left(d\phi-kA_1\right)\wedge\beta_3 \, +\, dA_1(x^\mu)\wedge \omega_2\, + \, \ldots
\end{equation}
and so the 4d Lagrangian (\ref{lagrangiano-Abeliano}) arises from the dimensional reduction of the 11d kinetic term $dA_3 \wedge *_{11} dA_3$. These observations will be exploited and generalized in the next subsections in the context of type IIB compactifications, in order to reproduce via dimensional reduction the 4d Lagrangian of non-Abelian discrete gauge symmetries. 

\subsection{Non-Abelian discrete symmetries from torsion homology}
\label{subsec:torsionforms}

Torsion classes have appeared in an example in \cite{Gukov:1998kn} as a source of discrete non-Abelian gauge symmetries in 5d in the AdS/CFT setup (see also \cite{Burrington:2006uu,Burrington:2007mj}). In this subsection we further explore and generalize this realization in the 4d setup, unveiling that the key to non-Abelianity lies in the existence of wedge (or cup) product relations among torsion classes. The corresponding dimensional reduction allows an elegant derivation of a general class of 4d theories with non-Abelian discrete gauge symmetry.
 
\subsubsection{Non-Abelian strings and the Hanany-Witten effect}

In order to describe the link between non-Abelian discrete gauge symmetries and torsion let us consider the class of models given by type IIB compactifications to 4d. In a generic 6d manifold there are two independent torsion classes, corresponding to torsion 1-cycles (and 4-cycles) and torsion 2-cycles (and 3-cycles)
\begin{equation}
\textrm{Tor }H_1(X_6,\IZ)\simeq \textrm{Tor }H_4(X_6,\IZ) \quad \textrm{and} \quad \textrm{Tor }H_2(X_6,\IZ)\simeq \textrm{Tor }H_3(X_6,\IZ)\ \label{torsion-homology}
\end{equation}
The first class actually describes two different kinds of discrete gauge symmetries: one of them associated to spontaneously broken U(1) symmetries that result from reducing the RR 2-form $C_2$  and the other to the spontaneously broken U(1)'s that result from reducing the NSNS 2-form $B_2$. In the latter case the 4d particles and strings charged under the discrete gauge symmetry arise from fundamental strings wrapping torsion 1-cycles and NSNS 5-branes wrapping torsion 4-cycles, respectively, while in the former case they arise from D1 and D5-branes. On the other hand, the second class in (\ref{torsion-homology}) describes discrete gauge symmetries associated to the RR 4-form $C_4$, with charged particles and strings arising from D3-branes wrapping torsion 3-cycles and 2-cycles, respectively.

As emphasized above, in compactifications with torsion classes the key to non-Abelianity is encoded in the existence of relations between torsion elements. Let us be more specific and consider the simple case where the torsion groups of $X_6$ are given by
\begin{equation}
\textrm{Tor\,} H_1(X_6,\IZ)=\textrm{Tor\,} H_4(X_6,\IZ)=\IZ_{k} \quad \ \textrm{Tor\,} H_2(X_6,\IZ)=\textrm{Tor\,} H_3(X_6,\IZ)=\IZ_{k'}\label{simplest}
\end{equation}
In general $k\neq k'$ although their precise relation is not relevant for our momentary purposes. Naively, considering general $(p,q)$-strings and 5-branes, the torsion 1-cycles would seem to produce a $\IZ_k\times \IZ_k$ symmetry, while also considering D3-branes in torsion cycles would add an extra $\IZ_{k'}$ factor. This mere Abelian structure is however promoted to a non-Abelian one if the corresponding classes have non-trivial relations. Indeed, if the torsion 4-cycles dual to the 1-cycles intersect non-trivially along a torsion 2-cycle, there is a non-trivial Hanany-Witten effect \cite{Hanany:1996ie} between the 4d strings obtained from NS5 and D5-branes wrapping the torsion 4-cycles. Crossing the strings in 4d leads to the creation of D3-branes wrapped on the torsion 2-cycle at the intersection of the 4-cycles, namely the creation of a 4d string associated to the RR 4-form. This 4d string creation effect is associated to non-Abelian discrete symmetry groups \cite{Alford:1989ch,Alford:1990mk,Alford:1990pt,
Alford:1991vr,Alford:1992yx,Lee:1994qg}. At the level of the gauge holonomies that result from moving around the 4d strings, we have the non-Abelian relation
\beqa
\tilde T_1\tilde T_2=\tilde T_3\tilde T_2\tilde T_1
\eeqa
among the generators $\tilde T_1$, $\tilde T_2$ of the two $\IZ_k$'s and the generator $\tilde T_3$ of $\IZ_{k'}$. This defines a finite Heisenberg group (c.f. eq.~(\ref{finite-heisenberg})). The same result can be obtained by working out the non-Abelian transformations undergone by particles moving around 4d strings, again by invoking the Hanany-Witten effect \cite{Gukov:1998kn}. 

\subsubsection{Dimensional reduction and four-dimensional effective action}

Just like in the Abelian case, this microscopic description of a non-Abelian discrete gauge symmetry should have a macroscopic counterpart via dimensional reduction. Indeed, we will show below how  a 4d effective Lagrangian reproducing such non-Abelian symmetries can be obtained by following the same procedure as in the Abelian case. Again, in order to perform the dimensional reduction we need to consider a set of non-harmonic forms satisfying (\ref{general-relations}), together with certain relations among them which are necessary for the non-Abelian pattern to emerge, and are equivalent to the topological conditions which allow for the Hanany-Witten effect. For simplicity, we will consider here the simple case where the torsion classes of $X_6$ are given by (\ref{simplest}). The more general case can be worked out in a similar way, as it is explicitly done in appendix \ref{app:general}.

More precisely, we consider a set of non-harmonic Laplacian eigenforms in  $X_6$ 
\begin{align}
d\gamma_1&=k\rho_2 \ , & d\tilde\rho_{4}&=k\zeta_{5}\label{formst}\\
d\alpha_3&=k'\tilde{\omega}_4 \ , & d\omega_{2}&=k'\beta_{3}\nonumber
\end{align}
with $\rho_2$, $\tilde{\omega}_4$, $\zeta_5$ and $\b_3$ representing the generators of the torsion cohomology Poincar\'e dual to (\ref{simplest}), and such that
\begin{equation}
\int_{X_6}\gamma_1\wedge \zeta_{5}=\int_{X_6}\rho_2\wedge \tilde\rho_{4}=\int_{X_6}\alpha_3\wedge \beta_{3}=\int_{X_6}\omega_{2}\wedge\tilde\omega_4=1
\end{equation}
In these expressions $k^{-1}$ and $k'^{-1}$ are the torsion linking numbers between dual $p$- and $(5-p)$-cycles, with $p=1,3$ respectively, and encode the monodromies which are felt by an electric (magnetic) charge when moved in a closed loop around its dual magnetic (electric) source. The fact that these torsion cycles have a non-trivial intersection pattern as described above is expressed in terms of these dual forms as
\begin{equation}
\rho_2\wedge\rho_2=M\, \tilde\omega_4
\label{intert}
\end{equation}
with $M\in\IZ$, which can be integrated to\footnote{In principle, instead of (\ref{idt}) one could have chosen the more general condition
\be
\nonumber
\rho_2\wedge \gamma_1=M'\, \alpha_3 + M'' \beta_3  \quad \quad M', M'' \in \IZ
\ee
This choice however, corresponds to gauging also the magnetic degrees of freedom and it will not be explored here.}
\begin{equation}
\rho_2\wedge \gamma_1=M'\, \alpha_3 \quad \quad M' \in \IZ \ \, {\rm such \ that}\ \, k \, M=k'\, M'
\label{idt}
\end{equation}

Let us then perform dimensional reduction of the type IIB supergravity action, taking into account the relations that we have introduced above. The relevant part of the action written in the 10d Einstein frame is
\begin{equation}
S_{\rm 10d}=\frac{1}{4\kappa_{10}^2}\int d^{10}x\left[(-G_E)^{1/2}\left(-\mathcal{M}_{ij}\, dB_2^i\cdot dB_2^j-\frac12 (F_5)^2\right)+\frac{\epsilon_{ij}}{2}dC_4\wedge B_2^i\wedge dB_2^j\right]\label{10d-actiont}
\end{equation}
where $B_2^1\equiv B_2$ and $B_2^2\equiv C_2$ are respectively the NSNS and RR 2-form potentials, $F_5=dC_4-C_2\wedge dB_2$, the matrix $\mathcal{M}_{ij}$ denotes the SO(2)$\backslash$SL(2,$\IZ$) coset metric
\begin{equation}
\mathcal{M}_{ij}=\frac{1}{\textrm{Im }\tau}\begin{pmatrix}|\tau|^2 & -\textrm{Re }\tau\\ -\textrm{Re }\tau& 1\end{pmatrix}
\end{equation}
and $\tau=C_0+ie^{-\phi}$ the complex axio-dilaton.

In order to dimensionally reduce this action, we expand the NSNS and RR 2-forms and the RR 4-form potentials as\footnote{This expansion is the most general one if we assume an underlying orientifold structure, according to which $\gamma_1$ and $\rho_2$ must be odd and $\omega_{2}$, $\tilde\omega_4$, $\alpha_3$ and $\beta_{3}$ even forms under the orientifold action. We also ignore 4d 2-forms resulting from $B_2^i$, as they do not play any role in what follows.}
\begin{align}
B_2^i&=b^{i}\rho_2+A_1^i\wedge \gamma_1\, \qquad i=1,2\label{b2t}\\
C_4&=b^3\tilde\omega_4+A_1^3\wedge\alpha_3+V_1^{3}\wedge\beta_{3}+c_2\wedge \omega_{2}\nn
\end{align}
obtaining several 4d vectors and scalars. The corresponding 10d field-strengths read
\begin{align}
\label{Fsi}
dB_2^i&=\eta^i\wedge\rho_2+dA^i_1\wedge \gamma_1\ , \qquad i=1,2\\
F_5&=\eta^3\wedge \tilde\omega_{4}
-F_{2}^3\wedge \alpha_3
+\tilde{F}_2^{3}\wedge \beta_{3}+dc_2\wedge \omega_{2}
\label{f5t}
\end{align}
where we have introduced the following 4d 1-form potentials
\begin{equation}
\eta^i_\mu\equiv \partial_\mu b^i-kA^i_\mu\ , \qquad
\eta^3_\mu\equiv
\partial_\mu b^3-k'A^3_\mu-M b^2\eta^1_\mu\label{physicalt}
\end{equation}
and field-strengths
\begin{equation}
k'F_{2}^3\equiv d\eta^3-\frac{\epsilon_{ij}}{2}M  \eta^i\wedge \eta^j\ , \qquad
\tilde{F}_2^{3}\equiv dV^{3}_1+k' c_2 \label{f2t}
\end{equation}
and we have made use of the relations (\ref{intert}) and (\ref{idt}).
Substituting these expansions into eq.~(\ref{10d-actiont}) we get (up to total derivatives and in 4d Planck mass units)
\begin{multline}
S_{\rm 4d}=\frac14\int d^{4}x\left[(-g)^{1/2}\left(-\mathcal{M}_{ij}\mathcal{N}dA^i_{1}\cdot dA^j_{1} -\mathcal{M}_{ij}\mathcal{T} \eta^i\cdot \eta^j-\frac{\mathcal{R}}{2} (F_{2}^3)^2+\right.\right.\\
\left.\left.+\mathcal{Q} F_{2}^3\cdot \tilde{F}_{2}^{3}+\frac{\mathcal{S}}{2} (\tilde{F}^{3}_2)^2-\frac{\mathcal{G}}{2}(dc_2)^2-\frac{\mathcal{G}^{-1}}{2} (\eta^3)^2\right)-\eta^0\wedge dc_2-\tilde{F}^{3}_2\wedge F_{2}^3\right]\label{4dintt}
\end{multline}
where we have defined\footnote{Note that idempotency of the hodge operator imply the non-trivial relation $\mathcal{R}\mathcal{S}+\mathcal{Q}^2=-1$, so these quantities are not all independent.}
\begin{align}
\mathcal{N}&\equiv \int_{{X}_6}\gamma_1\wedge *_6\gamma_1\ , & \mathcal{T}&\equiv \int_{{X}_6}\rho_2\wedge *_6\rho_2\ , \label{metricst}\\
\mathcal{Q}&\equiv \int_{{X}_6}\alpha_3\wedge *_6\beta_{3}\ , & \mathcal{R}&\equiv \int_{{X}_6}\alpha_3\wedge *_6\alpha_3\ ,\nonumber\\
\mathcal{S}&\equiv \int_{{X}_6}\beta_{3}\wedge *_6\beta_{3}\ , & \mathcal{G}&\equiv \int_{{X}_6}\omega_{2}\wedge *_6\omega_{2}\ ,\nonumber
\end{align}
Since we have not yet imposed the self-duality condition of the RR 5-form field-strength, $F_5=*_{10}F_5$, the 4d effective action (\ref{4dintt}) contains redundant degrees of freedom. Making use of
\begin{equation}
\tilde{F}^{3}_2=-F_{2}^3\mathcal{Q}\mathcal{S}^{-1}-*_4F_{2}^3\mathcal{S}^{-1}\ , \qquad
dc_2=\mathcal{G}^{-1}*_4\eta^3
\end{equation}
we finally obtain
\begin{multline}
S_{\rm 4d}=\frac14\int d^{4}x\left[(-g)^{1/2}\left( -\mathcal{M}_{ij}\mathcal{T} \eta^i\cdot \eta^j-\mathcal{G}^{-1}(\eta^3)^2\right.\right.\\
\left.\left.-\mathcal{M}_{ij}\mathcal{N}dA^i_{1}\cdot dA^j_{1}+\mathcal{S}^{-1}(F_{2}^3)^2\right)+\mathcal{Q}\mathcal{S}^{-1}F_{2}^3\wedge F_{2}^3\right]\label{4dfinalt}
\end{multline}
From the first line of this equation and comparing to  eq.~(\ref{metrica-gauged}) we observe that the 4d axion-like scalars in this setup span a gauged scalar manifold  with tangent space metric
\begin{equation}
\mathcal{P}_{ab}=-\frac14 \begin{pmatrix}
\mathcal{G}^{-1}& 0 \\
0&\mathcal{T}\mathcal{M}_{ij}
\end{pmatrix}\ ,
\end{equation}
right-invariant 1-forms given by the eqs.~(\ref{gauged-forms-example}) upon the following identifications
\begin{align}
&\phi^1=b^1\ , \qquad \phi^2=b^2\ , \qquad \phi^3=b^3-\frac{M}{2}b^1b^2\ , \qquad k_1=k_2=k\ , \qquad k_3=k'\label{identifications}\\
&\left. A^1_\mu\right|_{\rm sec.\ 3.3}=A^1_\mu\ , \qquad \left. A^2_\mu\right|_{\rm sec.\ 3.3}=A^2_\mu\ , \qquad \left. A^3_\mu\right|_{\rm sec.\ 3.3}=A^3_\mu-\frac{M'}{2}(b^1A^2_\mu+b^2 A^1_{\mu})\ ,\nn
\end{align}
and structure constants of the Heisenberg algebra $\mathcal{H}_3$. The example based on the Heisenberg manifold $\tilde{\mathcal{M}}=\mathcal{H}_3(\IR)/\Gamma$ discussed in section \ref{subsec:heisenberg-example} is thus physically realized in a large class of type IIB compactifications with torsional homology.

\subsubsection{Non-Abelian discrete gauge symmetries}

As the 4d effective action (\ref{4dfinalt}) is identical to the one analyzed in section \ref{subsec:heisenberg-example}, the discrete gauge symmetries that one obtains from it can be directly extracted from the discussion therein. It is however illustrative to reproduce the previous 4d discussion from a 10d perspective. In the present context, the shift symmetries of the scalars $b^1$, $b^2$, and $b^3$ are inherited from the 10d gauge transformations of $B_2$, $C_2$ and $C_4$. Indeed, at the perturbative level we have that  the 10d field strengths $dB_2$, $dC_2$ and $F_5$, eqs.~(\ref{Fsi}) are invariant under any of the following shifts
\begin{equation}
B_2\to B_2+ \eps^1 \rho_2\ , \qquad
C_2\to C_2+ \eps^2 \rho_2\ , \qquad C_4\to C_4+ \eps^2 \rho_2\wedge B_2+\eps^3 \tilde\omega_4\ ,
\end{equation}
with $\eps^{1,2,3} \in \IR$. Hence, they are symmetries of the Lagrangian (\ref{10d-actiont}). Upon dimensional reduction they become isometries of this axionic manifold, which at this level can be thought to be $\mathcal{H}_3$. On the other hand, one should impose the discrete identifications
\begin{align}
C_4&\to C_4 + \tilde\omega_4\label{c4gauget}\\
B_2&\to B_2+\rho_2\ , \nn \\
C_2&\to C_2+\rho_2\ , \qquad C_4\to C_4+\rho_2\wedge B_2\ ,\nn
\end{align}
which in 4d become the discrete transformations
\begin{align}
b^1&\to b^1 + 1\ ,\label{discrete-symmetryt}\\
b^2&\to b^2 + 1\ , \qquad b^3\to b^3+ M b^1\nn\\
b^3 &\to b^3 + 1\ , \nn
\end{align}
in agreement with eqs.~(\ref{discrete-example}) once we make use of the identifications (\ref{identifications}). These symmetries generate a non-Abelian discrete group $\Gamma$, so that the final axionic manifold is $\tilde{\CM} = \mathcal{H}_3/\G$. The corresponding algebra generators satisfy eq.~(\ref{alg}) and the symplectic Sp(2,$\, \IZ$) $\simeq$ SL(2,$\, \IZ$) global structure of this algebra is in this context inherited from the SL(2,$\, \IZ$) invariance of the 10d action.

Because of the torsion, the discrete shifts of $B_2$, $C_2$ and $C_4$ above not only imply the discrete transformations (\ref{discrete-symmetryt}), but also discrete transformations of the 4d massive gauge vectors $A^i$ that must occur simultaneously with them. That is, we find that the discrete shifts of the scalars are gauged to
\begin{align}
&A^1_\mu \to A^1_\mu + \partial_\mu\lambda^1 \ ,  \qquad \qquad A^2_\mu \to A^2_\mu + \partial_\mu\lambda^2\ ,  \label{discrete-gaut}\\
&
A^3_\mu \to A^3_\mu + \partial_\mu\lambda^3 +M' k\lambda^2A^1_\mu+M'b^1\partial_\mu \lambda^2\ ,\nn \\
&b^1\to b^1+k\lambda^1 \ , \qquad b^2\to b^2+k\lambda^2 \ ,\qquad
b^3 \to b^3 + Mk b^1\lambda^2+k'\lambda^3\ .\nn
\end{align}
Compatibility with the discrete transformations (\ref{discrete-symmetryt}) leads to a sublattice $\hat\Gamma\subset\Gamma$, as in eqs.~(\ref{discrete-gauge-example}). 

As already discussed, the gauge symmetries of the action (\ref{4dfinalt}) for non-vanishing $k$ and $k'$ are then given by the quotient ${\bf P}=\Gamma/\hat\Gamma$.
It is insightful to work out the transformation of charged fields under such discrete gauge group. For that aim, consider a 4d charged particle $\psi(x)$ with integer charges $q_I$ under $A^I_1$, $I=1,2,3$. From a 10d perspective this corresponds to a bound state of $q_0$ D3-branes wrapping the torsion 3-cycle above, and $q_1$ fundamental strings and $q_2$ D1-branes wrapping the torsion 1-cycle. The 4d covariant derivative is given by
\begin{equation}
D\psi(x)=\left[d+iq_I \hat{A}^I_1\right]\psi(x)\label{covariantt}
\end{equation}
with $\hat{A}^i_1=k^{-1}\eta^i$, $i=1,2$, and $\hat{A}^3_1=k'^{-1}\eta^3$. In general, under a discrete gauge transformation (\ref{discrete-gaut}) the field $\psi(x)$ will transform with a holonomy phase and a charge redefinition. Indeed, acting on (\ref{covariantt}) with (\ref{discrete-gaut}) we obtain the following transformation properties under the action of ${\bf P}$
\begin{align}
&\tilde T_1 \ : \quad \psi(x) \to \textrm{exp}\left[2\pi i k^{-1} q_1 \right]\psi(x)\\
&\tilde T_2 \ : \quad \psi(x) \to \textrm{exp}\left[2\pi i k^{-1} q_2 \right]\mathcal{U}\psi(x)\nn\\
&\tilde T_3 \ : \quad \psi(x) \to \textrm{exp}\left[2\pi i k'^{-1} q_3\nn \right]\psi(x)\label{chargedt}
\end{align}
where $\mathcal{U}$ is the charge redefinition
\begin{equation}
\mathcal{U}\ : \quad \begin{pmatrix}q_1\\ q_2\\ q_3\end{pmatrix}\ \to \ \begin{pmatrix}1&0&M'\\
0&1&0\\
0&0&1\end{pmatrix}\begin{pmatrix}q_1\\ q_2\\ q_3\end{pmatrix}
\end{equation}
The above monodromies can also be derived from a higher dimensional point of view, by simply performing the discrete shifts (\ref{discrete-symmetryt}) on the Chern-Simons actions of the corresponding type IIB $p$-branes, and reading the induced charges before and after the shift. Note that a particle with charge $q_I$ is indistinguishable from a particle with charge $q_i + k n^i$ (or $q_3 + k' n^3$ for the case $I=3$) for $n^i\in\IZ$, and therefore represent the same physical state, in agreement with the underlying discrete symmetry discussed above. Moreover, due to the non-Abelian structure the basis of charge eigenstates of $\tilde T_1$ and $\tilde T_2$ are not compatible with each other, and the two types of charges cannot be simultaneously measured.

\subsubsection{A simple example revisited}
\label{subsec:simple-example}

In order to illustrate the usefulness of the above results let us consider the simple setup of \cite{Gukov:1998kn}, consisting on a set of $N$ fractional D3-branes at a $\IC^3/\IZ_3$ singularity on type II string theory. In the large $N$ limit this setup backreacts to string theory on $AdS_5\times \IS^5/\IZ_3$, dual to a certain supersymmetric SU($N$)$^3$ gauge theory with bifundamental matter. The SCFT has a $\Delta(27)$ discrete symmetry, which can be obtained from torsion homology in the 5d AdS dual, as described in \cite{Gukov:1998kn}. Alternatively, we can make use of the results of the previous subsection for torsion $p$-forms in order to make explicit the non-Abelian discrete gauge symmetry directly from dimensional reduction of the backreacted setup.
Indeed, in this case the torsion homology of $\IS^5/\IZ_3\times \IS^1$ corresponds to eqs.~(\ref{simplest}) with $k=k'=3$ and $M=M'=1$ in eq.~(\ref{intert}). Charged particles in the 4d theory are thus labeled by three fractional charges $\frac{1}{N}(q_1,q_2,q_3)$, with $q_I$ defined mod $3$. 
In particular, the three types of bifundamental fields described in \cite{Gukov:1998kn} correspond to states $\psi_r(x)$, $r=1,\ldots,3$ with $(q_1,q_2,q_3)=(r-1,0,1)$. These are 4d particles which result from wrapping a D3-brane in the torsion 3-cycle and 0, 1 or 2 fundamental strings in the torsion 1-cycle. From our previous results, we obtain that the three generators of discrete symmetries act in these states as
\begin{align}
&\tilde T_1 \ : \quad (\psi_1, \psi_2, \psi_3)\ \to\ (\psi_1, \xi\psi_2, \xi^2\psi_3)\nonumber\\
&\tilde T_2 \ : \quad (\psi_1, \psi_2, \psi_3)\ \to\ (\psi_2, \psi_3, \psi_1)\\
&\tilde T_3 \ : \quad (\psi_1, \psi_2, \psi_3)\ \to\ (\xi\psi_1, \xi\psi_2, \xi\psi_3)\nonumber
\end{align}
with $\xi=e^{2\pi i/3N}$, in complete agreement with the results of \cite{Gukov:1998kn}.

\section{Non-Abelian discrete symmetries from discrete isometries}
\label{sec:isometries}
\label{subsec:twisted-torus}

In Kaluza-Klein compactification, isometries of the compactification manifold produce gauge symmetries in the lower dimensional theory. This is familiar for continuous isometries, but also holds for discrete isometries, suggesting a natural source for (possibly non-Abelian) discrete gauge symmetries. Although this mechanism is seemingly different from the description in terms of gaugings in section \ref{sec:nonAbelian}, in this section we focus on an illustrative example which nicely fits within this framework.

Prototypical examples of compactification spaces with discrete isometries are twisted tori. For simplicity we focus on the case of a twisted torus $(\IT^3)_M$ (where $M$ denotes the first Chern class of the $\IS^1$ fibration over the base $\IT^2$). 
This space and its symmetries can be neatly displayed by the following coset construction (see e.g. \cite{Kachru:2002sk,Marchesano:2006ns}). Consider the set ${\cal H}_3(\IR)$ of upper triangular matrices
\beqa
g(x,y,z)\, =\, \begin{pmatrix}
1 & x & z+\frac{xy}2 \cr 0 & 1 & y \cr 0 & 0 & 1
\end{pmatrix}\quad , \quad x,y,z\in\IR
\eeqa
which forms a non-compact Heisenberg group under multiplication 
\beqa
g(x,y,z)g(x',y',z')=g(x+x',y+y',z+z'+\frac 12(xy'-x'y))
\eeqa
A basis of e.g. right-invariant forms $\eta^x  = dx$, $\eta^y=dy$, $\eta^z=dz-\frac 12(ydx-xdy)$ allows the introduction of a metric $ds^2=(\eta^x)^2+(\eta^y)^2+(\eta^z)^2$ with an isometry group defined by right multiplication, and therefore given by ${\cal H}_3(\IR)$ itself. More precisely, we have that the Killing vectors of this metric are given by the left-invariant vectors of ${\cal H}_3(\IR)$, a simple basis for them being
\begin{subequations}
\begin{align}
\label{isoVx}
X_L^x\, =\, \p_x -{\textstyle\frac 12}y \p_z   \quad & \quad  \quad g(x,y,z)\to g(x+\lam_x,y,z-{\textstyle\frac 12} y \lam_x)\\
\label{isoVy}
X_L^y\, =\, \p_y +{\textstyle\frac 12}x \p_z   \quad & \quad  \quad g(x,y,z)\to g(x,y+\lam_y,z+{\textstyle\frac 12} x \lam_y)\\
\label{isoVz}
X_L^z\, =\, \p_z   \quad & \quad  \quad g(x,y,z)\to g(x,y,z+ \lam_z)
\end{align}
\end{subequations}
where we have also specified the continuous isometries generated upon exponentiation of such Lie algebra elements.

The twisted torus is obtained as a left coset $(\IT^3)_M={\cal H}_3(\IR)/{\cal H}_3(M)$ of the non-compact space ${\cal H}_3(\IR)$ by the infinite discrete subgroup $\hat\G={\cal H}_3(M)$ with elements of the form
\beqa
\begin{pmatrix}
1 & Mn_x & Mn_z \cr 0 & 1 & Mn_y \cr 0 & 0 & 1
\end{pmatrix}\quad , \quad n_x,n_y,n_z\in\IZ
\eeqa
In other words, by imposing the identifications
\beqa
g(x,y,z)\,\sim\, g(x+M,y,z-{\textstyle\frac M2} y)\,\sim\, g(x,y+M,z+{\textstyle\frac M2} x)\,\sim\, g(x,y,z+M)
\label{identT3}
\eeqa
As the metric is made of right-invariant forms, $(\IT^3)_M$ has a well-defined quotient metric. 
On the other hand, some of the isometries of the parent space ${\cal H}_3(M)$ are broken in $(\IT^3)_M$.
The quotient enjoys a continuous U(1) isometry along the $\IS^1$ fiber, generated by the invariant Killing vector $X_L^z = X_R^z = \partial_z$. However, the other two vectors $X_L^x$ and $X_L^y$ are not right-invariant, and so the corresponding continuous isometries disappear. Indeed, one can see that the action of $X_L^x$ and $X_L^y$ is in general different for different points of ${\cal H}_3(M)$ which are identified under (\ref{identT3}). For instance,
\begin{align}
e^{\lam_x X_L^x} & :   g(x,y,z)\to g(x+\lam_x,y,z-{\textstyle\frac 12} y \lam_x)\\
e^{\lam_x X_L^x} & :   g(x,y+M,z+{\textstyle\frac M2} x)\to g(x+\lam_x,y+M,z+{\textstyle\frac M2} x-{\textstyle\frac 12} (y+M) \lam_x)\nn\\
& \hspace*{4cm} \sim g(x+\lam_x,y,z-{\textstyle\frac 12} y \lam_x +M \lam_x)\nn
\end{align}
and so these two actions are the same only if $\lam_x \in \IZ$. A similar statement holds for the parameter $\lam_y$ in (\ref{isoVy}). Hence, one finds that the identifications (\ref{identT3}) break two of the continuous isometries of the parent ${\cal H}_3(\IR)$, preserving only the discrete order-$M$ actions generated by
\beqa
e^{X_L^x}\, :\, g(x,y,z)\to g(x+1,y,z-{\textstyle\frac 12} y)\quad ,\quad e^{X_L^y}\, :\, g(x,y,z)\to g(x,y+1,z+{\textstyle \frac 12} x)
\eeqa
Just like $X_L^x$ and $X_L^y$, these generators do not commute, but rather produce an element of the U(1) generated by $X_L^z$, and realize a discrete Heisenberg group ${\bf P}=H_M = {\cal H}_3(M=1)/{\cal H}_3(M)$. This discrete non-Abelian isometry group produces a discrete non-Abelian gauge symmetry $H_M$ in the lower-dimensional theory\footnote{Note that although the twisted torus geometry has torsion cycles, the discrete gauge symmetry from discrete isometries is associated to components of the metric, and not to $p$-forms reduced on torsion classes, in contrast with the previous section.}.

The above construction is a particular case of a more general setup (see e.g. \cite{Castellani:1983tb,Coquereaux:1988ne}). Given a non-compact group $G$, the metric constructed with right-invariant forms has $G$ itself as its isometry group (by right multiplication). In taking the coset $G/H$ by a subgroup $H$, some of these isometries may survive (in continuous or discrete versions). In general, $H$ is not a normal subgroup of $G$, so $G/H$ is {\em not} a group, and cannot be the isometry group. To identify the correct isometry group, note that a point $g_1$ in $G/H$ is, at the level of $G$, an equivalence class of points of the form $g_2=g_1\gamma$, with $\gamma\in H$. An isometry $R$ in $G$, mapping such $g_1$ and $g_2$ to $g_1R$ and $g_2R$, is an isometry in $G/H$ if the images are in the same equivalence class, namely if $g_2R=g_1R\gamma'$ for some $\gamma'\in H$. This requires $R$ to satisfy $R^{-1}\gamma R= \gamma'$, namely conjugation by $R$ should leave $H$ invariant (although not necessarily pointwise). Those transformations form the so-called normalizer group $N_H$ of $H$, and define the maximal subgroup of $G$ such that $H$ is normal in $N_H$. Since $H$   acts trivially on $G/H$, the actual isometry group of $G/H$ is $N_H/H$. 

It is easy to show that in the twisted torus the group $N_{{\cal H}_3(M)}/{\cal H}_3(M)$ corresponds to the one identified above, namely $H_M \times$ U(1). The simplicity of the twisted torus allows to explicitly compute interesting restrictions imposed by the discrete symmetry on couplings of the lower-dimensional theory, as analyzed in detail in Appendix \ref{subsec:yukawas-kk}.

It is natural to ask if, besides the above higher-dimensional description, there is a  lower-dimensional description of the discrete gauge symmetry in terms of gauging of suitable scalars. Indeed, it is familiar that compactification on a twisted torus can alternatively be viewed as a compactification on $\IT^3$ with metric fluxes, which can be described in terms of gauging a Heisenberg algebra \cite{Kaloper:1999yr}. The qualitative structure of the gauging is already manifest in the twisted torus metric, with $g_{xz}\sim y$ and $g_{yz}\sim x$, as follows.  A gauge transformation of the KK gauge boson $V^x_\mu\sim g^x{}_{\mu}$ along the circle parametrized by $y$ (i.e. a translation in $y$) shifts the vev of the scalar $\phi\sim g_{xz}$, and similarly for the KK gauge boson along $x$ and the scalar $g_{yz}$. The integer $M$ arises as the ratio of winding numbers of the map between full translations in the geometric circles, and the induced shifts in the scalar manifold. The non-Abelian structure of the isometries of the scalar manifold makes the resulting discrete gauge symmetry non-Abelian. This qualitative description can be fleshed out by performing the dimensional reduction explicitly; this is carried out in detail in the related but more interesting case of magnetized toroidal compactification in the next section.

\section{Magnetized branes and discrete flavour symmetries}
\label{sec:magnetized}

In this section we discuss the appearance of non-Abelian discrete symmetries in magnetized toroidal compactifications, focusing on magnetized D-brane systems, although similar conclusions hold for analogous heterotic models and T-dual intersecting brane models (for review of these constructions, see \cite{Ibanez:2012zz} and references therein). These symmetries are analogous to those in the twisted torus in the previous section, since dimensional reduction of the latter on the $\IS^1$ fiber produces a $\IT^2$ compactification with a constant magnetic field for the KK gauge boson. 
We start our analysis with the case of magnetized $\IT^2$, to make the main ideas manifest, and also to allow contact with the earlier geometric discussion for twisted tori; subsequently we move on and analyze the more involved system of magnetized $\IT^6$ compactifications. For the latter, and via dimensional reduction of the 10d type I supergravity action, we will make direct contact with the formalism of section \ref{sec:nonAbelian}.

\subsection{Non-Abelian discrete symmetries and Yukawa couplings in magnetized $\IT^2$}
\label{magnetized-toy}

As a warm up, let us consider a $\IT^2$ compactification with a U(1) gauge field background
\beqa
A_1\, =\, \pi M\, (xdy-ydx) \ , \quad {\rm so \ that\ } \; F_2= 2\pi M\, dx\wedge  dy
\label{gauge-field-def}
\eeqa
Before introducing $F_2$ the translations generated by $\p_x$ and $\p_y$ are clearly symmetries of the system. When introducing a non-vanishing $F_2$, even if constant along $\IT^2$, they are no longer so, since $A_1$ depends explicitly on its coordinates $x,y$
\bea
A_1(x+\lam_x,y) \, = \, A_1(x,y) + \lam_x d\chi_x & \quad \quad & \chi_x \, =\, \pi M y \\
A_1(x,y+\lam_y) \, = \, A_1(x,y) + \lam_y d\chi_y & \quad \quad & \chi_y \, =\, - \pi M x\nn
\eea
Hence, if we want to leave our system unchanged, with every translation we need to perform a gauge transformation that compensates the change in $A_1$. Acting on a wavefunction of charge $q$, this means that we need to perform the operations
\bea
\psi(x,y) & \raw & e^{-iq \lam_x \chi_x} \psi(x+\lam_x,y) \, =\, e^{q\lam_x  X_x} \psi(x,y)\\
\psi(x,y) & \raw & e^{-iq \lam_y\chi_y} \psi(x, y+\lam_y) \, =\, e^{q\lam_y  X_y} \psi(x,y)\nn
\eea
instead of plain translations. The above are generated by the operators $X_x$, $X_y$, defined as (we also introduce the generator of gauge transformations $X_Q$)
\beq
X_x= \, \partial_x -i\pi My\quad , \quad X_y=\partial_y+i\pi Mx\quad ,\quad X_Q=2\pi i
\eeq
These are the analogues of the left-invariant vectors of the twisted torus. Indeed, they satisfy the Heisenberg algebra $[X_x,X_y]=MX_Q$, which exponentiates to the group
\beqa
&& g(\epsilon_x,\epsilon_y,\epsilon_Q)=\exp \left(\frac{\epsilon_x}{M} X_x +\frac{\epsilon_y}{M} X_y +\frac{\epsilon_Q}{M} X_Q \right)\label{magnetized-group}\\
&& g(\epsilon_x',\epsilon_y',\epsilon_Q')g(\epsilon_x,\epsilon_y,\epsilon_Q)=g\left(\epsilon_x+\epsilon_x',\, \epsilon_y+\epsilon_y',\, \epsilon_Q+\epsilon_Q'+\frac{\epsilon_x'\epsilon_y}{2M}-\frac{\epsilon_x\epsilon_y'}{2M}\right)
\nn
\eeqa
Again, the  continuous version of this group is not a symmetry of our system. The point is that since the two-torus is compact, we need to impose well-defined boundary conditions on our charged particles, namely
\be
\psi(x+1,y)\,  =\, e^{iq \chi_x} \psi(x,y) \quad \quad {\rm and} \quad \quad \psi(x,y+1)\,  =\, e^{iq \chi_y} \psi(x,y)
\ee
In order to be actual symmetries of the system, the actions of $X_x$, $X_y$ and $X_Q$ must be compatible with the above identifications. This is automatic for $X_Q$, but not for $X_x$ and $X_y$, since
\be
e^{iq\lam_x X_x} \psi(x, y+1) \, =\, e^{iq\chi_y} e^{q\lam_x X_x} \psi(x, y) \quad \iff \quad e^{iq \lam_x M}\, =\, 1
\ee
which is only true if $\lam_x qM \in \IZ$. Similarly, we obtain that $\lam_y qM \in \IZ$ and so, for particles of minimal charge $q=1$ the symmetry corresponds only to a set of  discrete elements together with the gauge transformations generated by $X_Q$, namely
\beqa
{\bf P}=\{g(n_x,n_y,\epsilon_Q)\, |\, n_x,n_y=0,\ldots, M-1; \epsilon_Q\in\IR\}\, =\, {H}_M \times \textrm{U(1)}
\label{magnetized-group-discrete}
\eeqa

Notice that in order to arrive to the above conclusion it was not necessary to know the precise form of the wavefunctions in a magnetized torus. This is to be expected because (\ref{magnetized-group-discrete}) is a symmetry group of the background, and not of its fluctuations. Nevertheless such symmetry group should have a well-defined action on the magnetized torus wavefunctions, which should transform as a particular representation under the discrete group $H_M$. Indeed, by solving for the $q=1$ wavefunctions of a magnetized $\IT^2$ one finds (see, e.g., \cite{Cremades:2004wa})
\be
\psi^{j, M} (z, U) =  e^{i\pi M z \pim z/\pim U} 
\cdot 
\vartheta
\left[
\begin{array}{c}
\frac{j}{M} \\ 0
\end{array}
\right]
(M z, M U)
\label{totalsoln}
\ee
where $U$ stands for the complex structure and $z= x + U y$ the complex coordinate of the $\IT^2$,  $j \in \IZ\, {\rm mod\, } M$ is a family index  and $\vt$ is the Jacobi theta function
\be
\vt \left[
\begin{array}{c}
r \\ p
\end{array}
\right] (\nu,U) =  \sum_{l \in \IZ} 
e^{\pi i (r + l)^2 U} \ e^{2\pi i (r + l)(\nu + p)}
\label{theta}
\ee
One can now check that the action of the symmetry group (\ref{magnetized-group-discrete}) on this set is given by 
\begin{equation}
g(n_x,n_y,\epsilon_Q)\, \psi^{j, M} (z, U)\, =\, e^{2\pi i (\epsilon_Q+  n_xn_y/2M)} e^{2\pi i \frac{n_xj}{M}}\, \psi^{j+n_y,M} (z, U)
\end{equation}
with $n_x$, $n_y$ and $\epsilon_Q$ taken as in (\ref{magnetized-group-discrete}). Notice that acting on the vector of functions
\be
{\Psi}\, =\, 
\left(
\begin{array}{c}
\psi^{0,M} \\  \vdots \\ \psi^{M-1,M}
\end{array}
\right)
\ee
the action of $g$ preserves the norm $\sum_j |\psi^j|^2$ and corresponds to an element of U($M$). In particular, the discrete parameters $n_x$, $n_y$ that generate the group $H_M$ are mapped to the 't Hooft clock and shift $M\times M$ matrices
\beqa
{\bf P}(1,0,0)\, \raw\, \tilde T_x\equiv\begin{pmatrix} 1 &  & & \cr &  \omega & & \cr  & & \ldots & \cr & & & \omega^{M-1} \end{pmatrix}\quad \quad
{\bf P}(0,1,0)\, \raw \, \tilde T_y\equiv\begin{pmatrix}  & 1 & & \cr & & 1 & \cr & & & 1 \cr 1 & & & \end{pmatrix}
\label{clock-shift}
\eeqa
with $\omega$ the $M$-th root of unity. Hence, via its action on wavefunctions, the discrete gauge group $H_M$ is embedded into a non-Abelian discrete subgroup of SU($M$).

The above system can be equivalently described as gaugings of a $\IT^2$ compactification (see \cite{Kaloper:1999yr} for a heterotic description, and \cite{Aldazabal:2008zza} for a D-brane/F-theory setup). In fact, the gauging structure is already manifest in (\ref{gauge-field-def}), as follows.  A gauge transformation of the KK gauge boson $V^x_\mu\sim g^x{}_{\mu}$ along the circle parametrized by $y$ (i.e. a translation in $y$) shifts the vev of the Wilson line scalar $\xi_x\sim A_{x}$, and similarly for the KK gauge boson along $x$ and the Wilson line scalar along $y$. The integer $M$ arises as the ratio of winding numbers of the map between full translations in the geometric circles and the induced shifts in the Wilson line scalars. The non-Abelian structure is manifest in the above Heisenberg algebra, which corresponds to the gauging algebra (\ref{alg}). In this respect, the appearance of the discrete Heisenberg group gauge symmetry in the compactified theory fits within the general perspective in section \ref{sec:nonAbelian}. Note that such picture implies that performing translations along the coordinates $x$ and $y$ should be equivalent to performing shifts in the corresponding axion scalars, which for the gauging associated to magnetization are the $\IT^2$ Wilson lines. Indeed, in the presence of Wilson lines the wavefunction (\ref{totalsoln}) generalizes to
\be
\psi^{j, M} (z+\xi, U) =  e^{i\pi M (z+\xi) \pim (z+\xi)/\pim U} 
\cdot 
\vartheta
\left[
\begin{array}{c}
\frac{j}{M} \\ 0
\end{array}
\right]
(M (z+\xi), M U)
\label{wlsoln}
\ee
with $\xi=-\xi_y+U\xi_x$, and so a translation in $\IT^2$ can be traded for a change in the Wilson line, and viceversa, in agreement with the gauging picture. This qualitative description can be fleshed out by performing the dimensional reduction of the U(1) theory on a magnetized $\IT^2$, as we analyze  in the next section for the more complete case of magnetized $\IT^6$ compactifications. 

Before that, we pause to emphasize the effect of these non-Abelian discrete gauge symmetries at the level of the 4d effective action, in particular as selection rules for charged matter Yukawa couplings.\footnote{That Yukawas and other couplings are constrained by discrete gauge symmetries is not only true for magnetized D-brane models, but holds in general. For instance, one obtains selection rules on the Yukawas arising from twisted tori compactifications, as shown in Appendix \ref{subsec:yukawas-kk}.} For simplicity, we consider the case where all charged matter fields involved have equal range $M$, and transform under the discrete Heisenberg group with the clock and shift matrices (\ref{clock-shift}). Further possibilities, with different field multiplicities and transformations, are illustrated by the example in section \ref{subsec:mssm}. Hence our present case involves couplings
\beqa
\lambda_{ijk}\, \Phi_i^{ab}\Phi_j^{bc}\Phi_k^{ca}
\eeqa
where $i,j,k=1,\ldots, M$ are family indices, and $a,b,c$ are Chan-Paton gauge indices. Since the massless 4d fields $\Phi_i$ have an internal wavefunction (\ref{totalsoln}), they also transform with the matrices (\ref{clock-shift}). The constraints imposed by the symmetry are
\beqa
&& \lambda_{ijk}=0\quad {\rm if}\;\, i+j+k\neq 0\quad {\rm mod}\;\, M \nonumber \\
&& \lambda_{ijk}=\lambda_{i+1,j+1,k+1}
\label{selection-rules}
\eeqa
These selection rules were obtained by explicit computation in \cite{Cremades:2004wa,Cremades:2003qj} for magnetized and intersecting brane models, respectively; they were suspected to arise from a discrete symmetry in \cite{Abe:2009vi} (see also \cite{Abe:2010iv}). Our analysis shows that this is not an accidental symmetry but rather a discrete gauge symmetry present in the model.

\subsection{Dimensional reduction and non-Abelian discrete symmetries}
\label{sec:dim-red-t6}

Let us now generalize the above simple picture and consider $N$ magnetized D9-branes on a $\IT^6 = (\IT^2)_1 \times (\IT^2)_2 \times (\IT^2)_3$ orientifold compactification with O9 and O5-planes (the conclusions hold for any system leading to the same 4d theory, in particular T-duals with lower-dimensional intersecting/magnetized branes). The 10d effective action for this setup can be suitably described in terms of the type I supergravity action 
\begin{equation}
S_{\rm 10d}=\frac{1}{2\kappa^2}\int d^{10}x(-G)^{1/2}\left[e^{-2\phi}(R+4\partial_\mu\phi\partial^{\mu}\phi)-\frac14|\tilde{F}_3|^2-\frac14|\tilde{F}_7|^2-e^{-\phi}\textrm{Tr}(|F_2|^2)\right]\label{10d-action-typeI}
\end{equation}
where we have doubled the degrees of freedom of $\tilde F_3$ by introducing a dual 7-form field-strength $\tilde F_7=-*\tilde F_3$, with
\begin{equation}
\tilde{F}_3=dC_2-\omega_3 \ , \qquad \tilde{F}_7=dC_6-\frac{1}{12}\omega_7
\end{equation}
and $\omega_3$ and $\omega_7$ respectively the 3- and 7-dimensional Chern-Simons forms
\begin{align}
\omega_3&=\textrm{Tr}_V\left[A\wedge dA -\frac{2i}{3}A\wedge A\wedge A\right]\\
\omega_7&=\textrm{Tr}_V\left[A\wedge dA\wedge dA\wedge dA -\frac{4i}{3}A\wedge A\wedge A\wedge dA\wedge dA-\right.\nn\\
&\left.\qquad \quad -\frac65 A\wedge A\wedge A\wedge A\wedge A\wedge dA+\frac{4i}{7}A\wedge A\wedge A\wedge A\wedge A\wedge A\wedge A\right]\nn
\end{align}
In order to achieve a chiral 4d compactification we magnetize the D9-branes by considering a background for the Yang-Mills field strength $F_2$ of the form
\begin{equation}
\label{flux2}
F_2=\sum_{r=1}^3\frac{\pi i}{\textrm{Im }U^r}\begin{pmatrix}
\frac{m_a^r}{n_a^r}\mathbb{I}_{n^r_a} & & & \\
& \frac{m_b^r}{n_b^r}\mathbb{I}_{n^r_b} & & \\
& & \frac{m_c^r}{n_c^r}\mathbb{I}_{n^r_c} & \\
& & & \ddots
\end{pmatrix}dz^r\wedge d\bar z^r
\end{equation}
where $z^r=dx^r+U^rdy^r$ is the complexified coordinate of $(\IT^2)_r$, $U^r$ its complex structure and $n^r_\alpha, m^r_\alpha \in \IZ$ the D9-brane `magnetic numbers', with $N=\sum_\alpha n^1_\alpha n^2_\alpha n^3_\alpha$. 

Upon dimensional reduction, and focusing on `diagonal' geometric moduli, the 4d effective theory contains $7+3N$ complex scalars: 3 complex structure moduli $U^p$, 3 K\"ahler moduli $T^p$, 1 axio-dilaton $S$ and $3N$ complex Wilson lines $\xi^p_\alpha$, that can be defined as \cite{Antoniadis:1996vw, Aldazabal:1998mr}
\begin{equation}
T^p=\int_{(\IT^2)_p} C_2 + ie^{-\phi} J \ , \qquad
S=\int_{\IT^ 6}C_6 + ie^{-\phi}\textrm{Vol}_6\ , \qquad \xi^p_\alpha=-\xi_{\alpha,y}^p + U^p\xi^p_{\alpha,x}
\label{defax}
\end{equation}
with $J$ the K\"ahler form of $\IT^6$, and $\textrm{Vol}_6 = J^3/3!$ its volume form. The scalars $\xi_{\alpha,x}^p$ and $\xi_{\alpha,y}^p$ are the real Wilson lines along the two 1-cycles of $(\IT^2)_p$, with periodicity $[0,2/n_\alpha^r)$.\footnote{A different (yet common) convention in the literature for the normalization of the Wilson line scalars is such that $\xi_{\alpha,x}^p,\xi_{\alpha,y}^p$ lay on the interval $[0,1/n_\alpha^p)$.} 

There are in addition $6+N$ U(1) gauge bosons in the 4d effective theory: 6 U(1) gauge bosons coming from the isometries of the $\IT^6$, that we shall represent by $V^{x,p}_\mu$ and $V^{y,p}_\mu$, and $N$ U(1) gauge bosons from the Cartan generators of the D9-brane U($N$) gauge group, denoted by $A_\mu^\alpha$ in what follows.

The kinetic terms for the 4d scalars can be obtained by dimensionally reducing the 10d action (\ref{10d-action-typeI}) on the above background (see also \cite{Kaloper:1999yr}), resulting in\footnote{We have taken the magnetization to be actually along the vector representation of SO($2N$), so that sums over $\alpha$ in eq.~(\ref{4d-magnetico}) and following expressions do not run over the orientifold brane images.}
\begin{multline}
\mathcal{L}_{\rm 4d}=\frac{1}{(S-\bar S)^2}\left|D S-\frac12\sum_{p=1}^3\sum_\alpha c^p_\alpha \left(\xi^p_{x,\alpha}D \xi^p_{y,\alpha}-\xi^p_{y,\alpha}D \xi^p_{x,\alpha}\right)\right|^2+\\
+\sum_{p=1}^3\left[\frac{1}{(U^p-\bar U^p)^2}|\partial U^p|^2+\frac{1}{(T^p-\bar T^p)^2}\left|D T^p+\frac12\sum_\alpha c^0_\alpha \left(\xi^p_{x,\alpha}D \xi^p_{y,\alpha}-\xi^p_{y,\alpha}D \xi^p_{x,\alpha}\right)\right|^2+\right.\\
\left.+\frac{1}{U^p-\bar U^p}\sum_\alpha\frac{c^0_\alpha}{T^p-\bar T^p}\left|-D\xi^p_{y,\alpha}+U^pD\xi^p_{x,\alpha}\right|^2\right]\label{4d-magnetico}
\end{multline}
where we have defined the following covariant derivatives
\begin{align}
\label{covader}
D_\mu S &= \partial_\mu S + \sum_\alpha d^0_\alpha A^\alpha_\mu & D_\mu T^p &= \partial_\mu T^p -\sum_\alpha d_\alpha^p A_\mu^\alpha\\
D_\mu \xi_{x,\alpha}^p &= \partial_\mu \xi_{x,\alpha}^p + \frac{m^p_\alpha}{n^p_\alpha}V^{y,p}_\mu & D_\mu \xi_{y,\alpha}^p &= \partial_\mu \xi_{y,\alpha}^p - \frac{m^p_\alpha}{n^p_\alpha}V^{x,p}_\mu \nn
\end{align}
Notice that the coefficients of this expression
\begin{align}
c^0_\alpha&=n^1_\alpha n^2_\alpha n^3_\alpha \ , & c^1_\alpha&=n^1_\alpha m^2_\alpha m^3_\alpha \ , & c^2_\alpha&=m^1_\alpha n^2_\alpha m^3_\alpha \ , & c^3_\alpha&=m^1_\alpha m^2_\alpha n^3_\alpha\\
d^0_\alpha&=m^1_\alpha m^2_\alpha m^3_\alpha \ , & d^1_\alpha&=m^1_\alpha n^2_\alpha n^3_\alpha \ , & d^2_\alpha&=n^1_\alpha m^2_\alpha n^3_\alpha \ , & d^3_\alpha&=n^1_\alpha n^2_\alpha m^3_\alpha\nn
\end{align}
measure the D9-, D5-, D3/$\overline{\rm D3}$ and D7/$\overline{\rm D7}$-brane charges of our system, induced on the stack of $N$ D9-branes by the magnetization.

In order to make contact with our general discussion of section \ref{sec:nonAbelian}, let us analyze the symmetries of the axion-like scalars within (\ref{4d-magnetico}). Due to the shift symmetries of the RR potentials in 10d, the real scalars $\phi^0\equiv \textrm{Re }S$ and $\phi^r\equiv \textrm{Re }T^r$ behave as axions in the 4d effective theory with shift symmetries
\begin{equation}
\phi^P\raw \phi^P+\epsilon^P \qquad \qquad P=0,1,2,3
\end{equation}
and discrete identifications
\begin{equation}
\phi^P\simeq \phi^P+1 \qquad \qquad P=0,1,2,3\label{apdiscr}
\end{equation}
The same occurs for the Wilson line scalars $\xi_{\alpha,x}^r$ and $\xi_{\alpha,y}^r$, for whom 4d shift symmetries descend from 10d YM gauge invariance. Setting momentarily $d_\a^P=0$, we see that in order to have a symmetry of the action (\ref{4d-magnetico}) a shift in the Wilson lines should be accompanied with a shift in the above RR axions. More precisely we have that
\begin{align}
\xi_{\alpha,x}^p&\raw \xi_{\alpha,x}^p+\epsilon_{\a,x}^p \ ,  & \phi^0 &\raw \phi^0+\frac12 c^p_\alpha \xi^p_{y,\alpha}\epsilon_{\a,x}^p \ , &     \phi^p  &\raw \phi^p-\frac12c^0_\alpha\xi^p_{y,\alpha}\epsilon_{\a,x}^p \\
\xi_{\alpha,y}^p&\raw \xi_{\alpha,y}^p+ \epsilon_{\alpha,y}^p \ , & \phi^0 &\raw \phi^0-\frac12c^p_\alpha\xi^p_{x,\alpha} \epsilon_{\alpha,y}^p \ ,  &   \phi^p &\raw \phi^p+\frac12c^0_\alpha\xi^p_{x,\alpha} \epsilon_{\alpha,y}^p\nn
\end{align}
leave (\ref{4d-magnetico}) invariant. We thus have the discrete identifications
\bea
\xi_{\alpha,x}^p \simeq \xi_{\alpha,x}^p+ \frac{2}{n_\alpha^{p}} \ , & & \phi^0 \simeq \phi^0+c^p_\alpha \frac{\xi^p_{y,\alpha}}{n_\alpha^{p}} \ , \quad  \phi^p \simeq \phi^p-c^0_\alpha\frac{\xi^p_{y,\alpha}}{n_\alpha^{p}} \label{xiydiscr0}\\
\xi_{\alpha,y}^p\simeq \xi_{\alpha,y}^p+ \frac{2}{n_\alpha^{p}} \ , & & \phi^0 \simeq \phi^0-c^p_\alpha\frac{\xi^p_{x,\alpha}}{n_\alpha^{p}} \ , \quad  \phi^p \simeq \phi^p+c^0_\alpha\frac{\xi^p_{x,\alpha}}{n_\alpha^{p}}\nn
\eea

Switching the coefficients $d_\a^P$ back on, the action (\ref{4d-magnetico}) can be rewritten in the form (\ref{gauged}). In particular, it can be written as a gauged non-Abelian scalar manifold with action (\ref{metrica-gauged}), right-invariant 1-forms (c.f. eq.~(\ref{1forms-gauged}))
\begin{align}
\eta^{\phi^p}_\mu&=\, \partial_\mu \phi^p+\frac12\sum_\alpha\left(-2 d_\alpha^p A_\mu^ \alpha+c^0_\alpha\xi^p_{x,\alpha}\eta_\mu^{\xi^p_{y,\alpha}}-c^0_\alpha\xi^p_{y,\alpha}\eta_\mu^{\xi^p_{x,\alpha}}\right)\label{magnetic-1forms}\\
\eta^{\phi^0}_\mu &=\,  \partial_\mu \phi^0+\frac12\sum_\alpha\left[2d_\alpha^0A_\mu^\alpha-\sum_{p=1}^3\left(c^p_\alpha\xi^p_{x,\alpha}\eta_\mu^{\xi^p_{y,\alpha}}-c^p_\alpha\xi^p_{y,\alpha}\eta_\mu^{\xi^p_{x,\alpha}}\right)\right]\nn\\
\eta_\mu^{\xi^p_{x,\alpha}}&=\, \partial_\mu \xi_{x,\alpha}^p + \frac{m^p_\alpha}{n^p_\alpha}V^{y,p}_\mu\nn\\
\eta_\mu^{\xi^p_{y,\alpha}}&=\, \partial_\mu \xi_{y,\alpha}^p - \frac{m^p_\alpha}{n^p_\alpha}V^{x,p}_\mu\nn
\end{align}
tangent space metric
\begin{multline}
\mathcal{P}_{ab}=\begin{pmatrix}
\mathcal{P}_{\phi^p\phi^p}&0&0&0\\
0&\mathcal{P}_{\phi^0\phi^0}&0&0\\
0&0&\mathcal{P}_{\xi^p_{x,\alpha}\xi^p_{x,\alpha}}&\mathcal{P}_{\xi^p_{x,\alpha}\xi^p_{y,\alpha}}\\
0&0&\mathcal{P}_{\xi^p_{y,\alpha}\xi^p_{x,\alpha}}&\mathcal{P}_{\xi^p_{y,\alpha}\xi^p_{y,\alpha}}
\end{pmatrix}=\\
\begin{pmatrix}
\frac{1}{(T^p-\bar T^p)^2}&0&0&0\\
0&\frac{1}{(S-\bar S)^2}&0&0\\
0&0&\frac{c^0_\alpha |U^p|^2}{(U^p-\bar U^p)(T^p-\bar T^p)}&-\frac{c^0_\alpha (U^p+\bar U^p)}{(U^p-\bar U^p)(T^p-\bar T^p)}\\
0&0&-\frac{c^0_\alpha (U^p+\bar U^p)}{(U^p-\bar U^p)(T^p-\bar T^p)}&\frac{c^0_\alpha}{(U^p-\bar U^p)(T^p-\bar T^p)}
\end{pmatrix}
\end{multline} 
and algebra of shift symmetries
\begin{equation}
[t_{x^p_\alpha},t_{y^p_\alpha}]=c^0_\alpha t_{\phi^p}-c^p_{\alpha}t_{\phi^0}\label{magneticgroup}
\end{equation}
where $t_{x^p_\alpha}$, $t_{y^p_\alpha}$, $t_{\phi^p}$ and $t_{\phi^0}$ denote the generators of shifts of the axion-like scalars $\xi^p_{x,\alpha}$, $\xi^p_{y,\alpha}$, $\phi^p$ and $\phi^0$, respectively.

From these expressions we observe that the coefficients $c_\a^P$ (i.e., the D9- and D5-brane charges our our model) determine the structure constants of the non-Abelian algebra in the axionic manifold $(\phi^P, \xi_{x,\a}^p, \xi_{y,\a}^p)$. On the other hand the coefficients $d_\a^P$ (the D3/$\overline{\rm D3}$ and D7/$\overline{\rm D7}$ charges) specify the  set of D-brane U(1)'s that become massive and the embedding of their gauge lattice into the lattice of scalar shifts. Indeed, as one can check from (\ref{covader}), the linear combinations of D9-brane U(1) gauge symmetries
\begin{equation} 
Q^P=\sum_\alpha d^P_\alpha Q^\alpha, \qquad P=0,1,2,3
\end{equation}
are spontaneously broken to discrete gauge symmetries by eating the RR scalars $\phi^P$, as it is familiar from the generalized Green-Schwarz mechanism in magnetized D9-brane compactifications. Similarly, the U(1) Kaluza-Klein isometries $V^{x,p}_\mu$ and $V^{y,p}_\mu$ 
are spontaneously broken to discrete isometries by eating Wilson line scalars. 

From  (\ref{magnetic-1forms}) one can check that under the U(1) gauge transformations the above axion-like scalars shift according to

\begin{center}
\begin{tabular}{c|c|c}
$Q^\alpha$ & $X^p$ & $Y^p$\\
\hline
$A^\alpha_\mu\to A^\alpha_\mu+\partial_\mu\lambda^1$ & $V^{x,p}_\mu\to V^{x,p}_\mu+\partial_\mu\lambda^2$ & $V^{y,p}_\mu\to V^{y,p}_\mu+\partial_\mu\lambda^3$\\
$\phi^0\to \phi^0-d^0_\alpha \lambda^1$ & $\phi^0\to \phi^0-\sum_\alpha d^0_\alpha\xi^p_{x,\alpha}\lambda^2$ & $\phi^0\to \phi^0-\sum_\alpha d^0_\alpha\xi^p_{y,\alpha}\lambda^3$\\
$\phi^p\to \phi^p+d^p_\alpha \lambda^1$ & $\phi^p\to \phi^p+\sum_\alpha d^p_\alpha\xi^p_{x,\alpha}\lambda^2$ & $\phi^p\to \phi^p+\sum_\alpha d^p_\alpha\xi^p_{y,\alpha}\lambda^3$\\
& $\xi^p_{y,\alpha} \to \xi^p_{y,\alpha}+\frac{m_\alpha^p}{n_\alpha^p}\lambda^2$ & $\xi^p_{x,\alpha} \to \xi^p_{x,\alpha}-\frac{m_\alpha^p}{n_\alpha^p}\lambda^3$\\
& $A^\alpha_\mu \to A^\alpha_\mu+\xi^p_{x,\alpha}\partial_\mu\lambda^2$ & $A^\alpha_\mu \to A^\alpha_\mu+\xi^p_{y,\alpha}\partial_\mu\lambda^3$
\end{tabular}
\end{center}
\begin{equation}
\label{magnetic-transform}
\end{equation}
This in turn implies that these gauge generators satisfy the gauge algebra \cite{Aldazabal:2008zza}
\begin{equation}
[X^p,Y^p]=-\frac{m^p_\alpha}{n^p_\alpha}Q^\alpha
\label{algmaggaug}
\end{equation}
The discrete identifications (\ref{apdiscr}) and (\ref{xiydiscr0}) are mapped via the above shifts to the discrete gauge symmetry group of the theory, which can be embedded in the continuous Lie group that arises from (\ref{algmaggaug}). Rather than describing the most general case, in what follows we illustrate the type of discrete gauge symmetries that one may obtain by analyzing a semi-realistic example.

\subsection{An example: flavour symmetries in a MSSM-like model}
\label{subsec:mssm}

We can illustrate the application of the above general ideas by considering the MSSM-like model of \cite{Cremades:2002qm,Cremades:2003qj,Cremades:2004wa} and its global realization in terms of an orientifold of $\IT^ 6/(\IZ_2\times \IZ_2)$ \cite{Marchesano:2004yq}. The model consists of two stacks of magnetized D9-branes (stacks $a$ and $d$), and two stacks of D5-branes (stacks $b$ and $c$). The wrapping and magnetization numbers are summarized in table \ref{modelo-guay}.

\begin{table}[!ht]
\begin{center}
\begin{tabular}{|c|c|c|c|}
\hline
$N_\alpha$ & $(n^1_\alpha,m^1_\alpha)$ & $(n^2_\alpha,m^2_\alpha)$ & $(n^3_\alpha,m^3_\alpha)$ \\
\hline \hline
$N_a = 3$ & $(1,0)$ & $(3,1)$ & $(3,-1)$\\
\hline
$N_b = 1$ & $(0,1)$ & $(1,0)$ & $(0,-1)$\\
\hline
$N_c = 1$ & $(0,1)$ & $(0,-1)$ & $(1,0)$\\
\hline
$N_d = 1$ & $(1,0)$ & $(3,-1)$ & $(3,1)$\\
\hline
\end{tabular}
\caption{Wrapping and magnetization numbers of the T-dual model to that of \cite{Cremades:2002qm,Cremades:2003qj} with D5 and magnetized D9-branes.\label{modelo-guay}}
\end{center}
\end{table}

If brane $b$ is not on top of the orientifold plane, the gauge group is $SU(3)\times SU(2)_L \times U(1)_Y\times U(1)_{B-L}\times \IZ_3$.\footnote{At other particular points of the moduli space, the continuous part of the gauge group can be enhanced to the maximal $SU(4)\times SU(2)_L \times SU(2)_R$ gauge symmetry of this model. See \cite{Cremades:2003qj} for details.} The two U(1) factors are related to the diagonal U(1) generators of the three stacks $a$, $c$ and $d$ as 
\begin{equation}
Q_{Y}=\frac16(Q_a-3Q_c+3Q_d)\ , \qquad Q_{B-L}=\frac{Q_a}{3}+Q_d
\end{equation}
whereas the remaining orthogonal combination of U(1)'s
\begin{equation}
Q_{\IZ_3}=3Q_a-Q_d
\end{equation}
is anomalous and is spontaneously broken to a discrete $\IZ_3$ gauge symmetry \cite{BerasaluceGonzalez:2011wy}.\footnote{More precisely, the anomalous U(1) is broken to a $\IZ_9$ discrete gauge symmetry, but a $\IZ_3\subset \IZ_9$ subgroup actually corresponds to the center of $SU(3)$. Hence, the only non-trivial discrete symmetry is $\IZ_9/\IZ_3\simeq\IZ_3$.} Indeed, observe from table \ref{modelo-guay} that the  the magnetization on the D9-branes induce non-trivial D7/$\overline{\rm D7}$ charges
\begin{equation}
d^2_a=d^3_d=3\ , \qquad d^3_a=d^2_d=-3
\end{equation}
so that from eq.~(\ref{magnetic-1forms}) we observe that 3U(1)$_a-$U(1)$_d$ becomes massive by combining with the linear combination of RR axions $\phi^2-\phi^3$.

\begin{table}[!ht]
\begin{center}
\begin{tabular}{|c|c|c|c|c|c|}
\hline
Sector & Field & $SU(3)\times SU(2)_L$ & $Q_Y$ & $Q_{B-L}$ & $Q_{\IZ_3}$ \\
\hline
\hline
$ab$ & $Q_L$ & $3{\bf (3,2)}$ & 1/6 & 1/3 & 3 \\
\hline
$ac$ & $U_R$ & $3{\bf (\bar 3,1)}$ & -2/3 & -1/3 & -3\\
\hline
$ac^*$ & $D_R$ & $3{\bf (\bar 3,1)}$ & 1/3 & -1/3 & -3\\
\hline
$db$ & $L$ & $3{\bf (1,2)}$ & -1/2 & -1 & 1 \\
\hline
$dc$ & $N_R$ & $3{\bf (1,1)}$ & 0 & 1 & -1\\
\hline
$dc^*$ & $E_R$ & $3{\bf (1,1)}$ & 1 & 1 & -1\\
\hline
$bc$ & $H_u$ & ${\bf (1,2)}$ & 1/2 & 0 & 0\\
\hline
$bc$ & $H_d$ & ${\bf (1,\bar 2)}$ & -1/2 & 0 & 0\\
\hline
\end{tabular}
\caption{Chiral spectrum, Higgs sector and charges of the model in table \ref{modelo-guay}.\label{materia-mssm}}
\end{center}
\end{table}

The chiral spectrum of the model is summarized in table \ref{materia-mssm}, and is exactly that of the MSSM with three generations of quarks and leptons and one vector-like pair of Higgses. As it has been noticed in \cite{BerasaluceGonzalez:2011wy}, the $\IZ_3$ discrete gauge symmetry of this model is equivalent to baryon triality \cite{Ibanez:1991pr}, up to U(1)$_{B-L}$ and U(1)$_Y$ transformations. In particular dimension five proton decay operators $Q_LQ_LQ_LL$ and $U_RE_RU_RD_R$ vanish to all orders in perturbation theory and, according to the discussion in section \ref{sec:instantons}, also at the non-perturbative level.\footnote{Baryon or lepton violating operators with dimension less than five are forbidden in this model because of the continuous U(1)$_{B-L}$ gauge symmetry.}

Besides this $\IZ_3$ discrete gauge symmetry, there are additional discrete gauge symmetries in this model that come from the isometries of $\IT^2\times \IT^2 \times \IT^2$ and act non-trivially on the flavour indices of the MSSM fields. Indeed, following our discussion in the previous subsection, we observe that the four translational symmetries of the second and third 2-tori are gauged and spontaneously broken down to $\IZ_3$ discrete gauge symmetries. Together with the above flavour-universal discrete symmetry, these symmetries form a non-Abelian discrete gauge symmetry algebra
\begin{equation}
[X^2_{\IZ_3},Y^2_{\IZ_3}]=-[X^3_{\IZ_3},Y^3_{\IZ_3}]=-\frac{Q_{\IZ_3}}{3}+\ldots
\end{equation}
where the dots in the r.h.s denote possible additional continuous U(1) generators. The four discrete isometry generators act on the MSSM fields as
\begin{align}
e^{X^2_{\IZ_3}}\ : & \qquad X_R^k \to e^{-\frac{2\pi i k}{3}}X_R^k\label{mssm-transform}\\
e^{X^3_{\IZ_3}}\ : & \qquad X_L^k \to e^{\frac{2\pi i k}{3}}X_L^k \nn\\
e^{Y^2_{\IZ_3}}\ : & \qquad (X^1_R, X^2_R, X^3_R)\to (X^2_R, X^3_R, X^1_R)\nn\\
e^{Y^3_{\IZ_3}}\ : & \qquad (X_L^1, X_L^2, X_L^3)\to (X_L^3, X_L^1, X_L^2)\nn
\end{align}
where $k=1,2,3$ denotes the three generations of MSSM fields and $X_R$ and $X_L$ denote collectively the right-handed and the left-handed MSSM fields, respectively. The resulting finite discrete symmetry group can be thought as two copies of $\Delta(27)$ acting respectively on the left or the right-handed MSSM fields and sharing a common flavour-universal center that contains $Q_{\IZ_3}$. 

The most interesting implications of flavour symmetries are the constraints they impose on the flavour structure of the couplings and, more particularly, on Yukawa couplings. In order to describe the structure of Yukawa couplings imposed by the non-Abelian discrete symmetry in this particular model, let us write them schematically as
\begin{equation}
\sum_{i,j=1}^ 3Y_{ij} X^i_L X^j_R H \label{generic-yukawa}
\end{equation}
where $Y_{ij}$ are holomorphic functions of the complex structure and the complex Wilson line scalars. In general, under a discrete gauge transformation the MSSM fields transform as in (\ref{mssm-transform}), so that $Y_{ij}$ will also transform accordingly such that the sum  (\ref{generic-yukawa}) remains invariant under discrete symmetry transformations. This, together with the fact that $Y_{ij}$ are holomorphic functions, leads to a set of constraints on the structure of the couplings. For the particular case at hand, we find that 
\begin{align}
&\frac{Y_{11}}{Y_{21}}=\frac{Y_{12}}{Y_{22}}=\frac{Y_{13}}{Y_{23}}\ , & &\frac{Y_{21}}{Y_{31}}=\frac{Y_{22}}{Y_{32}}=\frac{Y_{23}}{Y_{33}}\ , & &\frac{Y_{31}}{Y_{11}}=\frac{Y_{32}}{Y_{12}}=\frac{Y_{33}}{Y_{13}}\label{relations-yukawas}\\
&\frac{Y_{11}}{Y_{12}}=\frac{Y_{21}}{Y_{22}}=\frac{Y_{31}}{Y_{32}}\ , & &\frac{Y_{12}}{Y_{13}}=\frac{Y_{22}}{Y_{23}}=\frac{Y_{32}}{Y_{33}}\ , & &\frac{Y_{13}}{Y_{11}}=\frac{Y_{23}}{Y_{21}}=\frac{Y_{33}}{Y_{31}}\nn
\end{align}
The details on the derivation of these relations can be found in appendix \ref{app:details}. These relations imply that Yukawa couplings in this model have the structure
\begin{equation}
(Y_{ij})=\begin{pmatrix}
a_1b_1 & a_1b_2 & a_1b_3\\
a_2b_1 & a_2b_2 & a_2b_3\\
a_3b_1 & a_3b_2 & a_3b_3 
\end{pmatrix}
\end{equation}
with $a_i$ and $b_i$, $i=1,2,3$, holomorphic functions of the moduli. Intuitively, the fields $X_L^i$ and $X_R^j$ in (\ref{generic-yukawa}) are triplets under two different non-Abelian factors (albeit with a common center) associated to two different internal $\IT^2$'s; their transformations must cancel against those of $Y_{ij}$, which must be made up of two objects $a_i$ and $b_j$, transforming as conjugate triplets under the two factors.

The above result is in agreement with what was found in \cite{Cremades:2003qj,Cremades:2004wa} from a direct computation and in particular implies that the Yukawa matrices of this model have rank one. As we have already mentioned, discrete gauge symmetries are exact symmetries of the theory so this rank one structure will be preserved in the complete non-perturbative formulation of the model. In particular, for this model the rank one texture should survive through the instanton effects mentioned in \cite{Abel:2006yk}. Indeed, as we discuss in detail later on, and in analogy with the Abelian case, non-perturbative effects will in general induce couplings that violate the underlying continuous symmetries, but are invariant under the discrete gauge symmetry.  This results in a very much constrained flavor structure also for those non-perturbative couplings.

\subsection{K\"ahler potential and holomorphic variables}
\label{sec:holomorphic}

In the previous subsection we have made use of the holomorphic dependence of superpotential Yukawa couplings on the complex structure and complexified Wilson lines in order to obtain the selection rules that the discrete gauge symmetry imposes on them. As we will see in the next section, holomorphicity of the superpotential is also a key ingredient in deriving analogous rules for non-perturbatively induced superpotential couplings. Note however that, while the complex structure and  Wilson lines transform holomorphically under the transformations (\ref{magnetic-transform}), the complex axio-dilaton and K\"ahler scalars defined in eq.(\ref{defax}) in general transform non-holomorphically. Thus, the latter are not the right variables in terms of which the superpotential and gauge kinetic functions are holomorphic quantities. 

A simple method to obtain the suitable variables consists on expressing the 4d effective action (\ref{4d-magnetico}) in terms of the second derivatives of a K\"ahler potential
\begin{equation}
\mathcal{L}_{\rm 4d}=-\sum_{i,j}K_{i\bar j}\partial M^i\partial \bar M^{j}
\end{equation}
Indeed, after some algebra we find that the following K\"ahler potential
\begin{multline}
K=-\sum_{p=1}^3\left[\textrm{log}(U^p-\bar U^p)+\textrm{log}\left(\hat T^p-\bar{\hat{T}}^p-\frac12\sum_\alpha c^0_\alpha\frac{(\xi^p_\alpha-\bar\xi^p_\alpha)^2}{U^p-\bar U^p}\right)\right]-\\
-\textrm{log}\left(\hat S-\bar{\hat{S}}+\frac12\sum_\alpha\sum_{p=1}^3 c^p_\alpha\frac{(\xi^p_\alpha-\bar\xi^p_\alpha)^2}{U^p-\bar U^p}\right)
\end{multline}
correctly reproduces eq.~(\ref{4d-magnetico}),\footnote{In fact, the above K\"ahler potential leads to an extra term in the kinetic term of the complex Wilson line scalars that is not present in eq.~(\ref{4d-magnetico})
\begin{equation*}
K_{\xi^p_\alpha\bar\xi^p_\alpha}=-\frac{1}{U^p-\bar U^p}\sum_\alpha\left(\frac{c^0_\alpha}{T^p-\bar T^p}-\frac{c^p_\alpha}{S-\bar S}\right)
\end{equation*}
This terms perfectly agrees with the CFT result obtained in \cite{Lust:2004cx}. From this point of view, this extra term comes from the $\textrm{Tr}(|F_2|^4)$ term that we have neglected in eq.~(\ref{10d-action-typeI}).} where the redefined fields $\hat S$ and $\hat T^p$ are given by\footnote{Similarly, matter fields are also redefined by the Wilson line scalars. This redefinition can be seen for instance from the perturbative Yukawa couplings \cite{Camara:2009uv}. The latter carry an exponential prefactor which depends non-holomorphically on the Wilson line scalars, and that it is absorbed into a redefinition of the bifundamental fields.}
\begin{equation}
\hat S=S-\frac12\sum_\alpha\sum_{p=1}^3c^p_\alpha\frac{\xi^p_\alpha\textrm{Im }\xi^p_\alpha}{\textrm{Im }U^p}\ , \qquad \hat T^p=T^p+\frac12\sum_\alpha c^0_\alpha\frac{\xi^p_\alpha\textrm{Im }\xi^p_\alpha}{\textrm{Im }U^p}
\label{defhol}
\end{equation}
In particular, the discrete identifications (\ref{apdiscr}) and (\ref{xiydiscr0})  in terms of these variables now correspond to the holomorphic identifications (see also \cite{Camara:2009uv})
\begin{align}
\hat S&\simeq \hat S+1\label{holomorphic-ident}\\
\hat T^p&\simeq \hat T^p+1\nn \\
\xi^p_\alpha&\simeq \xi^p_\alpha+\frac{2}{n_\alpha^{p}}\nn \\
\xi^p_\alpha&\simeq \xi^p_\alpha+\frac{2U^p}{n_\alpha^{p}}& \hat S&\simeq \hat S- \frac{c^p_\alpha}{n_\alpha^p}\left(2\xi^p_\alpha+\frac{U^p}{n_\alpha^p}\right) &\hat T^p&\simeq \hat T^p+ \frac{c^0_\alpha}{n_\alpha^p}\left(2\xi^p_\alpha+\frac{U^p}{n_\alpha^p}\right)\nn
\end{align}
We will make use of these holomorphic variables when discussing instanton effects in the next section.

\section{Instantons}
\label{sec:instantons}

As we have seen in the previous section, the presence of non-Abelian discrete gauge symmetries in D-brane and other string theory models directly constrain the structure of Yukawa couplings at the perturbative level. A natural question is then if such discrete symmetries also affect those couplings that are generated at the non-perturbative level, in particular by instanton effects in 4d chiral compactifications. The purpose of this section is to show that this is indeed the case, and that most of the intuition that holds for instanton effects in compactifications with discrete Abelian symmetries generalizes to the non-Abelian case. 

In order to do so, let us recall the structure of instanton induced couplings in 4d chiral D-brane models, which is typically of the form (c.f. (\ref{instanton-operator})) 
\be
\Phi_1\Phi_2\dots \Phi_N\, \ca\, e^{-S_{\rm inst.}}
\label{operator}
\ee
where  
\be
S_{\rm inst.}\, =\, 2\pi (g_s^{-1} V + i \phi)
\ee
is the complexification of the D-instanton volume and $\Phi_i$ are 4d chiral open string modes. Finally, the prefactor $\ca$ is a non-trivial function of the open and closed string moduli of the compactification, excluding those closed string moduli that enter into D-instanton actions $S_{\rm inst.}$. The open string operator $\Phi_1\Phi_2\dots \Phi_N$ is non-trivially charged under a U(1) gauge symmetry arising from a bulk D-brane, symmetry that becomes massive by eating the axionic closed string modulus $\phi$. Both this term and exp$(-S_{\rm inst.})$ are not invariant under such U(1) gauge transformations and the corresponding shift in $\phi$, but their product is, so that (\ref{operator}) is an allowed operator. In case that the massive U(1) symmetry is not totally broken but a $\IZ_k$ subgroup remains, then exp$(-S_{\rm inst.})$ is invariant under the action of such $\IZ_k$ subgroup, and so must be $\Phi_1\dots \Phi_N$, so that not all operators can be generated in the effective theory \cite{BerasaluceGonzalez:2011wy}.

Let us now turn to the non-Abelian case. As we have seen in section \ref{sec:magnetized} when considering discrete symmetries in D-brane models we may not only focus on axions $\phi$ arising from the closed string sector, but also on open string axions $\xi_\a$. Hence, in order to check the transformation properties of each of the factors in (\ref{operator}) under non-Abelian transformations we need to consider the prefactor $\ca$ and its dependence on those open string axions that enter into the definition of the non-Abelian symmetry. 

The prefactor $\ca$ is oftentimes difficult to obtain, but  it can be explicitly computed in examples like toroidal compactifications with magnetized and/or intersecting D-branes.\footnote{This also applies to elliptically fibered Calabi-Yau compactifications where the interaction between open string chiral fields is localized at the elliptic fiber \cite{Uranga:2002pg}.} For instance, let us consider two magnetized D9-branes on an orientifold of $\IT^6 = (\IT^2)_1 \times (\IT^2)_2 \times (\IT^2)_3$ with magnetic numbers $(n_a^r,m_a^r)$ and $(n_b^r,m_b^r)$ as in (\ref{flux2}), and an Euclidean D1-brane wrapping  $(\IT^2)_p$'s. If this E1-brane has the appropriate zero mode structure and assuming that $d^r_a = -d^r_b = N$, a superpotential coupling like (\ref{operator}) will be generated for the open string fields $\Phi^{ab}_i$  that transforms in the bifundamental of U(1)$_a \times $U(1)$_b$. More precisely we will have something of the form
\be
\label{operator2}
 e^{-S_{\rm inst.}} \, \sum_{\a} \Phi_{\a_1} \dots \Phi_{\a_N}\, \ca_{\a_1}^\prime \dots \ca_{\a_N}^\prime
\ee
where each of the factors $\ca_{\a_i}$ arises from a three-point function of open string chiral fields, namely two fermionic zero modes of the E1-brane and a 4d chiral multiplet $\Phi_{\a_i}^{ab}$.  Since $\IT^6$ is factorized, such three-point functions are given by the product of three functions of the form
\be
\ca^\prime_{\d_{ijk}}\, =\, 
e^{i\pi M \xi \pim \xi/\pim U} 
\cdot 
\vartheta
\left[
\begin{array}{c}
\delta_{ijk} \\ 0
\end{array}
\right]
(M \xi, M U)
\label{yukawins}
\ee
one for each factor $(\IT^2)_r$ $r=1,2,3$, where for simplicity we have omitted the label $r$ of the $\IT^2$ in all these quantities. Here $U=U^r$ is the complex structure modulus of such $\IT^2$ and $\xi$ is a linear combination of complex open string moduli in $(\IT^2)_r$. Namely,
\be
 M \xi\, =\, (I_{bc} \xi_a + I_{ca} \xi_b)/d	\quad\quad\quad M \, =\,  I_{ab} I_{bc} I_{ca}/d^2
\ee
with $\xi_\a = \xi_\a^r$ defined as in (\ref{defax}) and $I_{ab} = I_{ab}^r \equiv n_a^r m_b^r - n_b^r m_a^r$ the number of zero modes that arise in the sector $ab$ from $(\IT^2)_r$. Similarly, one can define $I_{ca}= I_{ca}^r$ and $I_{bc}= I_{bc}^r$ as the zero modes of the E1-brane charged under U(1)$_a$ and U(1)$_b$, respectively, arising from $(\IT^2)_r$. Finally, $d = {\rm g.c.d.}( I_{ab}, I_{bc}, I_{ca})$ and we have that
\be
\d_{ijk}\, =\, \frac{i}{I_{ab}} + \frac{j}{I_{ca}} + \frac{k}{I_{bc}}
\ee
where $i,j,k$ label the chiral zero modes at each D-brane sector. In particular, the index $i$ labels 4d chiral fields $\Phi_\a$ in (\ref{operator2}) and $j$, $k$ the charged zero modes of the E1-instanton that couple to them.

It is easy to see that (\ref{yukawins}) is not a holomorphic function of the open string moduli $\xi_\a$ of the compactification. However, one may absorb the non-holomorphic prefactor exp$(i \pi M \xi \pim \xi/ \pim U)$ into the definition of the instanton classical action $S_{\rm inst.}$ and the chiral fields $\Phi_\alpha$. Indeed, as first pointed out in \cite{Camara:2009uv}, the whole expression (\ref{operator2}) can be rewritten as
\be
\label{operator3}
 e^{-\hat{S}_{\rm inst.}} \, \sum_{\a} \hat\Phi_{\a_1} \dots \hat\Phi_{\a_N}\, \ca_{\a_1}\dots \ca_{\a_N}
\ee
where $\hat{S}_{\rm inst.}$ is a linear function of the holomorphic variables $\hat{S}$, $\hat{T}^r$ defined in (\ref{defhol}), $\hat\Phi_\alpha$ are the redefined 4d chiral fields of \cite{Camara:2009uv} and the prefactors $\ca_\a$ are now holomorphic functions of the moduli. In the example at hand we have that $\hat{S}_{\rm inst.} = \hat{T}^p$, and that (\ref{yukawins}) gets replaced by
\be
\ca_{\d_{ijk}}\, =\, 
\vartheta
\left[
\begin{array}{c}
\delta_{ijk} \\ 0
\end{array}
\right]
(M \xi, M U)
\label{yukawins2}
\ee

One can now check how the non-perturbative coupling transforms under the discrete gauge symmetry, and in particular under  discrete Wilson line shifts. On the one hand we have
\be
\xi \quad \to \quad \xi + \frac{1}{M} \quad \quad \quad \ca_{\d_{ijk}} \, \to \, \ca_{\d_{ijk}} \, e^{2\pi i \d_{ijk}}
\label{shift1}
\ee
If $I_{ab}=I_{bc}=I_{ca}=d$ then (\ref{shift1}) corresponds to the third identification in (\ref{holomorphic-ident}), under which the other holomorphic variables do not transform. In particular $\hat{S}_{\rm inst.}$ remains invariant and since the product of $\ca$'s saturates all possible values $j$, $k$ for the charged instanton zero modes we obtain the transformation
\be
e^{-\hat{S}_{\rm inst.}} \ca_{\a_1}\dots \ca_{\a_N} \quad \to \quad e^{-\hat{S}_{\rm inst.}} \ca_{\a_1}\dots \ca_{\a_N} e^{2\pi i \sum_i \frac{\a_i}{I_{ab}}}
\label{transf1}
\ee
which means this term is invariant only if the flavor indices $\a_i$ add up to a multiple of $I_{ab}$, or in other words if
\be
\sum_i \a_i \, =\,0\  {\rm mod\ } I_{ab}
\ee
in analogy with the selection rules for perturbative Yukawa couplings.

On the other hand we have the shift
\be
\xi \quad \to \quad \xi + \frac{U}{M} \quad \quad \quad \ca_{\d_{ijk}} \, \to \, \ca_{\d_{ijk}+1/M} \, e^{-\pi i U/M} e^{-2\pi i \xi}
\label{shift2}
\ee
that can be partially compensated by a simultaneous shift of the form
\be
\hat{S}_{\rm inst.} \quad \to \quad \hat{S}_{\rm inst.} + 2\xi + U
\ee
as follows from the last identification in (\ref{holomorphic-ident}). Hence the product in the right hand side of (\ref{transf1}) remains invariant under (\ref{shift2}) except for a shift in the zero mode indices $(i,j,k) \to (i+i_0, j+j_0, k+k_0)$ such that
\be
i_0 I_{bc}I_{ca} + j_0 I_{ab}I_{bc} + k_0 I_{ca}I_{ab} \, =\, d^2
\label{shift3}
\ee
which is always possible. Notice that $\ca_{\d_{ijk}}$ only depends on the value of the l.h.s. of (\ref{shift3}), so given a prefactor $\ca_{\a_i}$ in (\ref{operator3}) there is a unique image $\ca_{\a_i'}$ under the shift (\ref{shift3}). We then we have that the second transformation acts as
\be
e^{-\hat{S}_{\rm inst.}} \ca_{\a_1}\dots \ca_{\a_N} \quad \to \quad e^{-\hat{S}_{\rm inst.}} \ca_{\a_1'}\dots \ca_{\a_N'}
\label{transf2}
\ee
and as a permutation of the chiral fields $\Phi_{\a_i}$ and instanton zero modes. Hence, if the operator (\ref{operator3}) is not invariant under this shift, the whole instanton amplitude should be a sum of operators of this form invariant  under (\ref{transf2}). 

\subsubsection*{An example}

Let us consider an example used in section 5 of \cite{Cremades:2003qj}, namely the case where there is only one $\IT^2$ and $I_{ab} = I_{bc} = I_{ca} =3$. There we have that
\bea
\ca_{\d_{111}} & = & 
\vartheta
\left[
\begin{array}{c}
0 \\ 0
\end{array}
\right]
(3 \xi, 3 U)\, =\,  \ca_{\d_{222}}\, =\,  \ca_{\d_{333}} \, \equiv\, A\\
\ca_{\d_{132}} & = & 
\vartheta
\left[
\begin{array}{c}
1/3 \\ 0
\end{array}
\right]
(3 \xi, 3 U)\, =\,  \ca_{\d_{213}}\, =\,  \ca_{\d_{321}} \, \equiv\, B\nn\\ 
\ca_{\d_{123}} & = &
\vartheta
\left[
\begin{array}{c}
-1/3 \\ 0
\end{array}
\right]
(3 \xi, 3 U)\, =\,  \ca_{\d_{231}}\, =\,  \ca_{\d_{312}} \, \equiv\, C\nn
\eea
all the other couplings vanishing. This induces a coupling of the form 
\be
e^{-\hat{S}_{\rm inst.}} \left[ABC\, \left(\hat\Phi_1^3 + \hat\Phi_2^3 + \hat\Phi_3^3\right) + (A^3 + B^3 + C^3) \hat\Phi_1\hat\Phi_2\hat\Phi_3 \right]  
\label{supoex}
\ee
which is indeed invariant under the discrete shifts (\ref{shift1}) and (\ref{shift2}), acting as
\bea
\label{shift1dex}
\xi & \to & \xi + \frac{1}{3} \quad \quad \quad A \to \,  A\ , \quad B \to \, e^{2\pi i/3} B\ , \quad C \to \, e^{-2\pi i/3} C  \\ 
\xi & \to & \xi + \frac{U}{3} \quad \quad \quad  A \, \to \, B  \, \to \, C \, \to \, A\nn
\label{shift2dex}
\eea
Notice however that none of the terms of (\ref{supoex}) is invariant individually.
Interestingly, for $\xi=0$ we have that $A=0$ and $B=-C$, so (\ref{supoex}) vanishes identically at that point.

\section{Conclusions}
\label{sec:conclusions}

In this paper we have analyzed the realization of non-Abelian discrete gauge symmetries in field theory and 4d string compactifications, and provided a fairly general `macroscopic' formulation based on the interplay between gauged non-Abelian isometries of the scalar manifold and non-trivial field identifications. We have studied several mechanisms to generate non-Abelian discrete gauge symmetries, and shown that they indeed fall into this formulation. 

In particular we have extended the realization in \cite{Gukov:1998kn,Camara:2011jg} of discrete gauge symmetries from NSNS and RR $p$-form fields in compactifications with torsion homology. We have also touched upon the discrete symmetries arising from discrete isometries of the compactification manifold, exemplified by twisted tori compactifications. The realization of the latter as a toroidal compactification with geometric fluxes suggests the extension to the study of discrete symmetries in other fluxed compactifications, which we leave for future work.  It would be interesting to generalize this analysis to other geometries, in particular CY spaces with (Abelian or non-Abelian) discrete isometries. 

Finally we have described non-Abelian discrete gauge symmetries in systems of magnetized gauge fields in toroidal compactifications (or quotients thereof). Although we focused on magnetized D-brane models, the results apply to analogous heterotic or T-dual intersecting brane models. We have derived the symmetry microscopically (from analysis of charged matter wavefunctions, and from dimensional reduction) and also shown its relation with the macroscopic gauging formulation. The discrete groups typically have a Heisenberg-like structure, with generators associated to discrete isometries of the torus geometry, commuting to discrete symmetries generated by the D-brane U(1)'s (broken to discrete subgroups as in \cite{BerasaluceGonzalez:2011wy}). We have shown that these symmetries imply powerful selection rules on the Yukawa couplings of charge matter fields, including those observed in \cite{Cremades:2003qj,Cremades:2004wa} (and their interpretation in  \cite{Abe:2009vi,Abe:2010iv}), and the rank-one structure in certain MSSM-like models \cite{Cremades:2003qj}, which are thus exact even at the non-perturbative level. It would be interesting to apply our insights to discrete symmetries in other semi-realistic constructions, e.g. to complement the recent discussions of discrete symmetries in heterotic orbifolds (see \cite{Kobayashi:2006wq,Nilles:2012cy} and references therein). 

As a final remark, note that our analysis is independent of the supersymmetry of the models. It would be interesting to particularize onto supersymmetric compactifications, and understand possible special properties of R-symmetries. We hope to come back to these and other appealing properties of discrete gauge symmetries in the future.

\bigskip

\centerline{\bf \large Acknowledgments}

\bigskip

We thank I. Garc\'{\i}a-Etxebarria and L.E. Ib\'a\~nez for useful discussions.  
This work has been partially supported by the grants FPA2009-09017, FPA2009-07908, FPA2010-20807 and Consolider-CPAN (CSD2007-00042)  from the Spanish Ministry of Economy and Competitiveness, HEPHACOS-S2009/ESP1473 from the C.A. de Madrid, AGAUR 2009-SGR-168 from the Generalitat de Catalunya and the contract ``UNILHC" PITN-GA-2009-237920 of the European Commission. M.B-G. acknowledges the finantial support of the FPU grant AP2009-0327. F.M. is supported by the Ram\'on y Cajal programme through the grant RYC-2009-05096 and by the People Programme of FP7 (Marie Curie Auction) through the REA grant agreement PCIG10-GA-2011-304023. D.R. is supported through the FPU grant AP2010-5687.

\newpage

\appendix

\section{Non-Abelian discrete symmetries from torsion forms: general case}
\label{app:general}

We have explored in subsection \ref{subsec:torsionforms} the structure of non-Abelian discrete gauge symmetries from the perspective of dimensional reduction, for the simplest case with torsion groups (\ref{simplest}). In this appendix we perform dimensional reduction of the type IIB action for the more general case with arbitrary torsion homology groups (\ref{torsion-homology}). Thus, we introduce  a set of non-harmonic eigenforms associated to the generators of the torsion homology groups with
\begin{align}
d\gamma_1^\alpha&=k^\alpha{}_\beta\rho_2^\beta \ , & d\tilde\rho_{4,\beta}&=k^\alpha{}_\beta\zeta_{5,\alpha}\label{forms}\\
d\alpha_3^\alpha&=k'^\alpha{}_\beta\tilde{\omega}_4^\beta \ , & d\omega_{2,\beta}&=k'^\alpha{}_\beta\beta_{3,\alpha}\nonumber
\end{align}
and
\begin{equation}
\int_{X_6}\gamma_1^\alpha\wedge \zeta_{5,\beta}=\int_{X_6}\rho_2^\alpha\wedge \tilde\rho_{4,\beta}=\int_{X_6}\alpha_3^\alpha\wedge \beta_{3,\beta}=\int_{X_6}\omega_{2,\beta}\wedge\tilde\omega_4^\alpha=\delta_\beta^\alpha
\end{equation}
In these expressions $k^{-1}$ and $k'^{-1}$ are the linking matrices between dual $p$- and $(5-p)$-cycles, with $p=1,3$ respectively.

We recast the torsion cycle intersection pattern in terms of these dual forms as
\begin{equation}
\gamma_1^\alpha\wedge \gamma_1^\beta=0
\ , \qquad \rho_2^\alpha\wedge \gamma_1^\beta=\mathcal{A}^{\alpha\beta}{}_\gamma\, \alpha_3^\gamma\ , \qquad \rho_2^\alpha\wedge\rho_2^\beta=\mathcal{K}^{\alpha\beta}{}_\gamma\, \tilde\omega_4^\gamma
\label{inter}
\end{equation}
where consistency with the exterior derivative requires
\begin{equation}
\mathcal{A}^{\alpha[\beta}{}_\gamma\, k^{\delta]}{}_\alpha=0\ , \qquad 
k^\alpha{}_{\beta} \, \mathcal{K}^{\delta\beta}{}_\gamma=k'^\beta{}_\gamma\, \mathcal{A}^{\delta\alpha}{}_\beta
\label{id}
\end{equation}

We proceed now to perform dimensional reduction of the type IIB supergravity action eq.~(\ref{10d-actiont}), taking into account these relations. Following the same reasoning than in subsection \ref{subsec:torsionforms}, we expand the NSNS and RR 2-forms and the RR 4-form as
\begin{align}
B_2^i&=b_{\alpha}^{i}\rho_2^\alpha+A^i_{1,\alpha}\wedge \gamma_1^\alpha\, \qquad i=1,2\label{b2}\\
C_4&=b^3_{\alpha}\tilde\omega_4^\alpha+A^3_{1,\alpha}\wedge\alpha_3^\alpha+V_1^{3,\alpha}\wedge\beta_{3,\alpha}+c_2^\alpha\wedge \omega_{2,\alpha}\nn
\end{align}
The corresponding 10d field-strengths read
\begin{align}
dB_2^i&=\eta_{\alpha}^{i}\wedge\rho_2^\alpha+dA^i_{1,\alpha}\wedge \gamma_1^\alpha\ , \qquad i=1,2\label{f5}\\
F_5&=\eta^3_{\beta}\wedge \tilde\omega_{4}^\gamma
-F_{2,\alpha}^3\wedge \alpha_3^\alpha
+\tilde{F}_2^{3,\alpha}\wedge \beta_{3,\alpha}+dc_2^{\alpha}\wedge \omega_{2,\alpha}\nn
\end{align}
where now
\begin{equation}
\eta^i_\alpha\equiv db^i_\alpha-k^\beta{}_\alpha A^i_{1,\beta}\label{physical}\ , \qquad
\eta^3_\alpha \equiv db^3_\alpha-k'^\beta{}_\alpha A^3_{1,\beta}-\mathcal{K}^{\gamma\rho}{}_\alpha b^2_\gamma\eta^1_\rho
\end{equation}
and
\begin{equation}
k'^\alpha{}_\beta F_{2,\alpha}^3\equiv d\eta^3_\beta-\frac{\epsilon_{ij}}{2}\mathcal{K}^{\rho\alpha}{}_\beta   \eta^i_\alpha\wedge \eta^j_\rho\label{f2}\ , \qquad
\tilde{F}_2^{3,\alpha}\equiv dV_1^{3,\alpha}+k'^\alpha{}_\beta c_2^\beta
\end{equation}
Substituting into eq.~(\ref{10d-actiont}) and making use of the relations (\ref{id}) we get
\begin{multline}
S_{\rm 4d}=\frac14\int d^{4}x\left[(-g)^{1/2}\left(-\mathcal{M}_{ij}\mathcal{N}^{\alpha\beta}dA^i_{1,\alpha}\cdot dA^j_{1,\beta} -\mathcal{M}_{ij}\mathcal{T}^{\alpha\beta} \eta^i_{\alpha}\cdot \eta^j_{\beta}-\right.\right.\\
\left.\left.-\frac{\mathcal{R}^{\alpha\beta}}{2} F_{2,\alpha}^3\cdot F_{2,\beta}^3+\mathcal{Q}^{\alpha}{}_{\beta} F_{2,\alpha}^3\cdot \tilde{F}_{2}^{3,\beta}+\frac{\mathcal{S}_{\alpha\beta}}{2} \tilde{F}^{3,\alpha}_2\cdot \tilde{F}^{3,\beta}_2-\frac{\mathcal{G}_{\alpha\beta}}{2}dc_2^\alpha\cdot dc_2^\beta-\right.\right.\\
\left.\left.-\frac{(\mathcal{G}^{-1})^{\alpha\beta}}{2} \eta_{\alpha}^3\cdot \eta_{\beta}^3\right)-\eta_{\alpha}^3\wedge dc_2^\alpha-\tilde{F}^{3,\alpha}_2\wedge F_{2,\alpha}^3\right]\label{4dint}
\end{multline}
where 
\begin{align}
\mathcal{N}^{\alpha\beta}&\equiv \int_{X_6}\gamma^\alpha_1\wedge *_6\gamma^\beta_1\ , & \mathcal{T}^{\alpha\beta}&\equiv \int_{X_6}\rho^\alpha_2\wedge *_6\rho^\beta_2\ , \label{metrics}\\
\mathcal{Q}^\alpha{}_\beta&\equiv \int_{X_6}\alpha_3^\alpha\wedge *_6\beta_{3,\beta}\ , & \mathcal{R}^{\alpha\beta}&\equiv \int_{X_6}\alpha_3^\alpha\wedge *_6\alpha_3^\beta\ ,\nonumber\\
\mathcal{S}_{\alpha\beta}&\equiv \int_{X_6}\beta_{3,\alpha}\wedge *_6\beta_{3,\beta}\ , & \mathcal{G}_{\alpha\beta}&\equiv \int_{X_6}\omega_{2,\alpha}\wedge *_6\omega_{2,\beta}\ ,\nonumber
\end{align}
and
\begin{equation}
\mathcal{R}^{\alpha\beta}\mathcal{S}_{\beta\gamma}+\mathcal{Q}^\alpha{}_\beta\mathcal{Q}^\beta{}_\gamma=-\delta^\alpha_\gamma\ , \qquad \mathcal{S}_{\alpha\beta}\mathcal{Q}^\beta{}_\gamma-\mathcal{Q}^\beta{}_\alpha\mathcal{S}_{\beta\gamma}=0
\end{equation}
The self-duality condition of the RR 5-form field-strength, $F_5=*_{10}F_5$, implies
\begin{equation}
\tilde{F}^{3,\alpha}_2=-F_{2,\beta}^3\mathcal{Q}^\beta{}_\gamma(\mathcal{S}^{-1})^{\gamma\alpha}-*_4F_{2,\beta}^3(\mathcal{S}^{-1})^{\beta\alpha}\ , \qquad
dc_2^\alpha=(\mathcal{G}^{-1})^{\alpha\beta}*_4\eta^3_{\beta}
\end{equation}
so we finally obtain
\begin{multline}
S_{\rm 4d}=\frac14\int d^{4}x\left[(-g)^{1/2}\left( -\mathcal{M}_{ij}\mathcal{T}^{\alpha\beta} \eta^i_{\alpha}\cdot \eta^j_{\beta}-(\mathcal{G}^{-1})^{\alpha\beta} \eta_{\alpha}^3\cdot \eta_{\beta}^3\right.\right.\\
\left.\left.-\mathcal{M}_{ij}\mathcal{N}^{\alpha\beta}F^i_{2,\alpha}\cdot F^j_{2,\beta}+(\mathcal{S}^{-1})^{\alpha\beta}F_{2,\alpha}^3\cdot F_{2,\beta}^3\right)+\mathcal{Q}^\alpha{}_\gamma(\mathcal{S}^{-1})^{\gamma\beta}F_{2,\alpha}^3\wedge F_{2,\beta}^3\right]\label{4dfinal}
\end{multline}

The gauge symmetries of this effective action are 
\begin{align}
&A^1_{1,\alpha} \to A^1_{1,\alpha} + d\lambda^1_\alpha\ , \qquad A^1_{2,\alpha} \to A^2_{1,\alpha} + d\lambda^2_\alpha \label{discrete-gau}\\
&A^3_{1,\alpha} \to A^3_{1,\alpha} + d\lambda^3_\alpha + \mathcal{A}^{\delta\beta}{}_\alpha k^\gamma{}_\delta (A^1_{1,\beta}\lambda^2_\gamma + b^1_\delta d\lambda^2_\beta)\nn\\
&b^1_\beta\to 
b^1_\beta + k^\alpha{}_\beta \lambda^1_\alpha\ , \qquad b^2_\beta\to b^2_\beta + k^\alpha{}_\beta \lambda^2_\alpha\ , \qquad b^3_\alpha\to b^3_\alpha+ \mathcal{K}^{\beta\delta}{}_\alpha k^\alpha{}_\beta \lambda^2_\alpha b^1_\delta+k'^{\alpha}{}_\gamma\lambda^3_\alpha \nonumber
\end{align}
These correspond to a set of non-commuting discrete $\IZ_{r^I_\beta}$ gauge symmetries, where $r^I_\beta$ is the lower integer for which $(k^{-1})_\alpha{}^\beta r^i_\beta$ (or $(k'^{-1})_\alpha{}^\beta r^3_\beta$ in the case of $I=3$) is an integer.

We can also work out the transformation of charged fields under these discrete gauge transformations. For that aim, we consider a 4d charged particle $\psi(x)$ with integer charges $q_I^\alpha$. The 4d covariant derivative is given by
\begin{equation}
D\psi(x)=\left[d+iq^\alpha_I \hat{A}^I_{1,\alpha}\right]\psi(x)\label{covariant}
\end{equation}
with $\hat{A}^i_{1,\alpha}=(k^{-1})_\alpha{}^\beta\eta^i_\beta$, $i=1,2$ and $\hat{A}^3_{1,\alpha}=(k'^{-1})_\alpha{}^\beta\eta^3_\beta$. Acting on (\ref{covariant}) with (\ref{discrete-gau}) we obtain the following transformation properties under the discrete gauge symmetry generators
\begin{align}
&\tilde T_1^\gamma \ : \quad \psi(x) \to \textrm{exp}\left[2\pi i (k^{-1})_\delta{}^\gamma q_1^\delta \right]\psi(x)\label{charged}\\
&\tilde T_2^\gamma \ : \quad \psi(x) \to \textrm{exp}\left[2\pi i (k^{-1})_\delta{}^\gamma q_2^\delta \right]\mathcal{U}\psi(x)\nn\\
&\tilde T_3^\gamma \ : \quad \psi(x) \to \textrm{exp}\left[2\pi i (k'^{-1})_\delta{}^\gamma q_3^\delta \right]\psi(x)\nn
\end{align}
where $\mathcal{U}$ is the charge redefinition
\begin{equation}
\mathcal{U}\ : \quad \begin{pmatrix}q_1^\alpha\\ q_2^\alpha\\ q_3^\alpha\end{pmatrix}\ \to \ \begin{pmatrix}\delta^\alpha_\beta&0&\mathcal{A}^{\gamma\alpha}{}_\beta\\
0&\delta^\alpha_\beta&0\\
0&0&\delta^\alpha_\beta\end{pmatrix}\begin{pmatrix}q_1^\beta\\ q_2^\beta\\ q_3^\beta\end{pmatrix}
\end{equation}

 \section{KK modes and Yukawas in twisted tori}
\label{subsec:yukawas-kk}

We have seen in section \ref{sec:isometries} that non-Abelian discrete isometries of the twisted torus $(\IT^2)_M=\mathcal{H}_3(\IR)/\mathcal{H}_3(M)$ lead to non-Abelian discrete gauge symmetries in the compactified effective theory. Thus, and in analogy to what occurs for Yukawa couplings in magnetized branes, we expect the presence of powerful selection rules in this setup for the couplings of KK modes. In this appendix we work out such selection rules for the three-point couplings, this time exploiting the underlying group structure of the twisted torus.

In more general terms, for a compactification on a group manifold $G/\Gamma$, where $G$ is a Lie group and $\Gamma\subset G$ a cocompact lattice, we expect 4d KK particles to arrange in irreducible unitary representations of the discrete isometry group ${\bf P}$ of $G/\Gamma$. Such representations can be explicitly worked out from the  irreducible representations of $G$ that are invariant under $\Gamma$. In physical terms, the components of these (generically infinite dimensional) representations correspond to wavefunctions of the particles in the 4d theory. The Clebsch-Gordan decomposition of the tensor product of two representations (namely, the operator product expansion, OPE) then allows the computation of superpotential couplings in the 4d effective theory, relating overlaps of $n$ wavefunctions to overlaps of two wavefunctions. Since the $\Gamma$-invariant irreducible representations of $G$ are also arranged in finite dimensional irreducible representations of the discrete symmetry group ${\bf P}$, the OPE must satisfy the set of selection rules associated to the discrete charge conservation.

In what follows we illustrate this procedure with the twisted torus compactification of section \ref{subsec:twisted-torus}, for which $G=\mathcal{H}_3(\IR)$ is the Heisenberg group.

\subsection{KK wavefunctions in twisted tori}

The irreducible unitary representations of the Heisenberg group can be worked out starting from eq.~(\ref{group}), for instance by means of Kirillov's orbit method (see e.g. Appendix D of \cite{Camara:2009xy} for details). In general, irreducible representations $\pi(g)$ of non-Abelian groups are not simple functions, but rather operators acting on a Hilbert space of functions $u(\vec s)\in L^2(\IR^{p(\pi)})$ with $p(\pi)\in{\bf N}$. For the case of the 3-dimensional Heisenberg group the complete set of irreducible unitary representations is given by
\begin{align}
\pi_k(\vec \phi)u(s)&=\textrm{exp}\left[2\pi ik\left(\phi^3+\frac{M}{2}\phi^1\phi^2+\phi^2 s\right)\right] u(s+M\phi_1)\ , \qquad u(s)\in L^2(\IR)\\
\pi_{k_1,k_2}(\vec\phi)&=\textrm{exp}\left[2\pi i\left(k_1\phi^1+k_2\phi^2\right)\right]\nn
\end{align}
$\Gamma$-invariant irreducible representations can be constructed by taking sums over the lattice $\Gamma$
\begin{equation}
B(g)\equiv \sum_{\gamma\in\Gamma}\pi(\gamma g)u(s)\label{bg}
\end{equation}
For the particular case of the Heisenberg group the complete procedure was carried out in \cite{Camara:2009xy}. Taking complex coordinates, $z=\phi^1+U\phi^2$, and imposing $B(g)$ to be eigenstates of the Laplacian (namely, of the quadratic Casimir invariant of $\mathcal{H}_3(\IR)$), we obtain
\begin{align}
B_{k,n,\delta}^M(z,\phi^3)&=\Psi^{kM}_{n,\delta}(z)\, \textrm{exp}\left(2\pi i k \phi^3\right)\label{landau}\\
B_{\mathbf{k}}(z)&=\textrm{exp}\left[2\pi i\frac{\textrm{Im}(\mathbf{k}z)}{\textrm{Im}(U)}\right]\nn
\end{align}
where $\mathbf{k}\equiv -k_2+\bar U k_1$, with $k_{1,2}\in {\bf N}$. We have defined
\begin{multline}
\Psi^N_{n,\delta}(z)\equiv\\
\left(2\pi|N|\right)^{\frac14}\sum_{s\in \IZ}\psi_n\left[\sqrt{2\pi|N|}\left(\frac{\delta}{N}+s+\frac{\textrm{Im}(z)}{\textrm{Im}(U)}\right)\right]\textrm{exp}\left[2\pi iN\textrm{Re}(z)\left(\frac{\delta}{N} + s + \frac{\textrm{Im}(z)}{\textrm{Im}(U)}\right)\right]
\end{multline}
with $n\in{\bf N}$, $\delta\in\IZ_N$ and $\psi_n(x)$ the Hermite functions given by
\begin{equation}
\psi_n(x)\equiv \frac{1}{\sqrt{n!2^n\pi^{1/2}}}H_n(x)e^{-x^2/2}
\end{equation}
where $H_n(x)$ are the standard Hermite polynomials.

\subsection{Yukawa couplings for KK modes}

We are particularly interested in 4d particles with wavefunctions of the type (\ref{landau}) as those carry a non-zero KK momentum along the fiber of the twisted torus and therefore see the non-Abelian nature of the gauging. In the language of magnetized D-branes these correspond to particles with non-trivial charge $k$ under the gauge symmetry of the D-brane. In such magnetized brane language, $n$ denotes the Landau level and $\delta$ runs over the degeneracy of the corresponding Landau level (namely, it is a flavour index).

For any given set of states (\ref{landau}) with fixed $k$ and $\delta$ (namely, for any given $\Gamma$-invariant representation of $G$ with non-vanishing central charge) the ground state $n=0$ (i.e., the highest weight of the representation) can be expressed in terms of Jacobi theta functions as
\begin{equation}
B^M_{k,0,\delta}(z,\phi^3)=(2\pi|kM|)^{\frac14}\, \vartheta\left[{\frac{\delta}{kM}\atop 0}\right](kMz;\, kMU)\, \textrm{exp}\left[i\pi kM\frac{z\textrm{Im}(z)}{\textrm{Im}(U)}+2\pi ik\phi^3\right]\label{ground}
\end{equation}
One may easily check that (\ref{ground}) transforms under the generators of the gauge lattice $\hat\Gamma$ as
\begin{align}
\phi^1&\to\phi^1+\frac{1}{M} \ , & \phi^3 &\to \phi^3 - \frac{\phi^2}{2}\ , &  B^M_{k,0,\delta} &\to \omega^\delta B^M_{k,0,\delta} \\
\phi^2&\to\phi^2+\frac{1}{M} \ , & \phi^3 &\to \phi^3 + \frac{\phi^1}{2}\ , & B^M_{k,0,\delta}&\to B^M_{k,0,\delta+k} \nn\\
\phi^3&\to\phi^3+\frac{1}{M}\ ,& & & B^M_{k,0,\delta}&\to \omega^k B^M_{k,0,\delta}\nn
\end{align}
with $\omega\equiv \textrm{exp}(2\pi i/M)$. As we saw in section \ref{subsec:twisted-torus}, these are the generators of the discrete gauge symmetry ${\bf P}=\Gamma/\Gamma'$ for the level $k$. For instance, for $k=1$ we have ${\bf P}= (\IZ_M\times \IZ_M)\rtimes \IZ_M$, and in the particular case of $M=3$, ${\bf P}=\Delta(27)$ as we have seen in different contexts in the main part of the text.

Let us now focus on the OPE of irreducible representations. For for zero-th Landau levels (\ref{ground}) the OPE can be easily worked out from the following relation between theta functions \cite{Cremades:2004wa}
\begin{multline}
\vartheta\left[{\frac{\delta_1}{N_1} \atop 0}\right](N_1z_1;\, N_1U)\, \vartheta\left[{\frac{\delta_2}{N_2} \atop 0}\right](N_2z_2;\, N_2U)=\\
\sum_{m\in\IZ_{N_1+N_2}}\vartheta\left[{\frac{\delta_1+\delta_2+N_1m}{N_1+N_2}}\right](N_1z_1+N_2z_2;\, (N_1+N_2)U)\cdot\, \\
\cdot \vartheta\left[{\frac{N_2\delta_1-N_1\delta_2+N_1N_2m}{N_1N_2(N_1+N_2)}}\right](N_1N_2(z_1-z_2);\, N_1N_2(N_1+N_2)U)
\end{multline}
which leads to,
\begin{multline}
\Psi^{N_1}_{0,\delta_1}(z_1)\Psi^{N_2}_{0,\delta_2}(z_2)\ \ \\ =\sum_{m\in\IZ_{N_1+N_2}}\Psi^{N_1+N_2}_{0,\delta_1+\delta_2+N_1m}\left(\frac{N_1z_1+N_2z_2}{N_1+N_2}\right)\Psi^{N_1N_2(N_1+N_2)}_{0,N_2\delta_1-N_1\delta_2+N_1N_2m}\left(\frac{z_1-z_2}{N_1+N_2}\right)\label{prod1}
\end{multline}
Setting $z_1=z_2=z$, $M_1=M_2=M$ and multiplying in both sides of this equation by $e^{2\pi i (k+j)\phi^3}$ we obtain
\begin{equation}
B^{M}_{k,\, 0,\, \delta_1}(z,\phi^3)\, B^{M}_{j,\, 0,\, \delta_2}(z,\phi^3)\  =\sum_{m\in\IZ_{M(k+j)}}B^{M}_{k+j,\, 0,\, \delta_1+\delta_2+kmM}(z,\phi^3)\, \Psi^{kj(k+j)M^3}_{0,\, M(j\delta_1-k\delta_2+kjmM)}(0)\label{ope1a}
\end{equation}
This OPE can be used to compute superpotential couplings which only involve 4d KK modes with zero-th Landau level. For instance, we see from the above expansion that, up to an overall normalization factor, 3-particle couplings are given by
\begin{equation}
Y_{(k,0,\delta_1)(j,0,\delta_2)(h,0,\delta_3)}\simeq \Psi^{kjhM^3}_{0,\, M(j\delta_3-h\delta_2)}(0)
\end{equation}
together with the selection rules,
\begin{equation}
h= k+j\ , \qquad
\frac{\delta_3-\delta_1-\delta_2}{kM}\in \IZ_{hM}\label{sela}
\end{equation}
Let us now generalize eq.~(\ref{ope1a}) to KK particles with higher Landau level. The key observation is that higher Landau levels can be obtained by acting with the creation operators on the lowest Landau level (namely, by acting with the lowering operator on the highest weight of the corresponding irreducible representation). The Heisenberg algebra has only one creation operator. This is given by
\begin{equation}
a^\dagger\equiv 2\frac{\partial}{\partial z}- \pi N\bar{z}
\end{equation}
Indeed, from eq.~(\ref{landau}) one may check that
\begin{equation}
a^\dagger\, \Psi^N_{n,\delta}=i\sqrt{4\pi |N|(n+1)}\, \Psi^N_{n+1,\delta}
\end{equation}
Acting  with this operator an arbitrary number of times on both sides of eq.~(\ref{prod1}) and performing some algebra, we obtain
\begin{multline}
\Psi^{N_1}_{n,\delta_1}(z_1)\Psi^{N_2}_{p,\delta_2}(z_2)=\sqrt{\frac{(-1)^{n+p}}{(N_1+N_2)^{n+p+1}}}\sum_{m\in\IZ_{N_1+N_2}}\sum_{\ell=0}^n\sum_{s=0}^p (-1)^s\sqrt{N_1^{n+s-\ell}N_2^{p+\ell-s}}\cdot\\ 
\cdot\left[\begin{pmatrix}
n\\ \ell
\end{pmatrix}\begin{pmatrix}
p\\ s
\end{pmatrix}\begin{pmatrix}
n+p-\ell-s\\
n-\ell
\end{pmatrix}\begin{pmatrix}
\ell + s\\
\ell
\end{pmatrix}\right]^{\frac12}\cdot\\
\cdot\Psi^{N_1+N_2}_{n+p-\ell-s,\, \delta_1+\delta_2+N_1m}\left(\frac{N_1z_1+N_2z_2}{N_1+N_2}\right)\Psi^{N_1N_2(N_1+N_2)}_{\ell+s,\, N_2\delta_1-N_1\delta_2+N_1N_2m}\left(\frac{z_1-z_2}{N_1+N_2}\right)\label{prod1a}
\end{multline}
Setting $z_1=z_2=z$, $M_1=M_2=M$ and multiplying in both sides of the equation by $e^{2\pi i (k+j)\phi^3}$), as we did before, we obtain the OPE for the complete set of KK modes of the 4d theory
\begin{multline}
B^{M}_{k,\, n,\, \delta_1}(z,\phi^3)\, B^{M}_{j,\, p,\, \delta_2}(z,\phi^3)=\sqrt{\frac{(-1)^{n+p}}{(k+j)^{n+p+1}M}}\sum_{m\in\IZ_{M(k+j)}}\sum_{\ell=0}^n\sum_{s=0}^p (-1)^s\sqrt{k^{n+s-\ell}j^{p+\ell-s}}\cdot\\ 
\cdot\left[\begin{pmatrix}
n\\ \ell
\end{pmatrix}\begin{pmatrix}
p\\ s
\end{pmatrix}\begin{pmatrix}
n+p-\ell-s\\
n-\ell
\end{pmatrix}\begin{pmatrix}
\ell + s\\
\ell
\end{pmatrix}\right]^{\frac12}\cdot\\
\cdot B^{M}_{k+j,\, n+p-\ell-s,\, \delta_1+\delta_2+kmM}(z,\phi^3)\, \Psi^{kjM^3(k+j)}_{\ell+s,\, M(j\delta_1-k\delta_2+kjmM)}(0)
\end{multline}
From this expression we easily read the 3-particle couplings for arbitrary KK modes in the 4d theory. Up to combinatorial and overall numeric factors, these are given by
\begin{equation}
Y_{(k,n,\delta_1)(j,p,\delta_2)(h,q,\delta_3)}\sim \Psi^{kjhM^3}_{n+p-q,\, M(j\delta_3-h\delta_2)}(0)
\end{equation}
together with the selection rules (\ref{sela}) and 
\begin{equation}
q\in \{0,1,\ldots, n+p\}
\end{equation}

\section{Details on the derivation of (\ref{relations-yukawas})}
\label{app:details}

In order to obtain the set of constraints that the discrete symmetry imposes on the holomorphic Yukawa couplings $Y_{ij}$ it is important to recall that if $f(\xi)$ is a holomorphic function of $\xi$ with domain on $\IC$ and is invariant under some discrete lattice $\Gamma$ then $f(\xi)$ is actually independent of $\xi$. Knowing how $Y_{ij}$ transform under the isometry generators, we can then build holomorphic invariants of $\{X^2_{\IZ_3},Y^2_{\IZ_3}\}$ and/or $\{X^3_{\IZ_3},Y^3_{\IZ_3}\}$ that are independent of the corresponding complex Wilson line scalars and that satisfy particular relations. 

Let us first illustrate the procedure on a similar model with only two generations of fields transforming as
\begin{align}
X^2_{\IZ_3}\ : & \qquad X_R^k \to e^{-i\pi k}X_R^k & Y^2_{\IZ_3}\ : & \qquad (X^1_R, X^2_R)\to (X^2_R, X^1_R)\\
X^3_{\IZ_3}\ : & \qquad X_L^k \to e^{i\pi k}X_L^k &
Y^3_{\IZ_3}\ : & \qquad (X_L^1, X_L^2)\to (X_L^2, X_L^1)\nn
\end{align}
with $k=1,2$. Yukawa couplings are of the form
\begin{equation}
Y=\begin{pmatrix}
Y_{11}&Y_{12}\\
Y_{21}&Y_{22}
\end{pmatrix}
\end{equation}
Taking into account the above transformations of the fields, we observe that the following function
\begin{equation}
A\equiv\frac{Y_{11}}{Y_{21}}-\frac{Y_{12}}{Y_{22}}
\end{equation}
is invariant under $X^3_{\IZ_3}$ and $(Y^3_{\IZ_3})^2$, so that $a$ is independent of the complex Wilson line scalar $\xi^3$. Moreover, under $Y^3_{\IZ_3}$ it transforms as
\begin{equation}
A\to -A
\end{equation}
but since acting with $Y^3_{\IZ_3}$ is equivalent to performing a shift in $\xi^3$,  this means that $a$ has to be identically zero. We have therefore shown that
\begin{equation}
\frac{Y_{11}}{Y_{21}}=\frac{Y_{12}}{Y_{22}}
\end{equation}

For the three generation model of section \ref{subsec:mssm} the proof follows the same logic. For instance, let us consider the following functions
\begin{equation}
A\equiv -\frac{Y_{11}}{Y_{21}}+\frac{Y_{12}}{Y_{22}}+\frac{Y_{13}}{Y_{23}}\ , \quad B\equiv \frac{Y_{11}}{Y_{21}}-\frac{Y_{12}}{Y_{22}}+\frac{Y_{13}}{Y_{23}}\ , \quad C\equiv \frac{Y_{11}}{Y_{21}}+\frac{Y_{12}}{Y_{22}}-\frac{Y_{13}}{Y_{23}}
\end{equation}
invariant under $X^3_{\IZ_3}$ and $(Y^3_{\IZ_3})^3$ and therefore independent of the complex Wilson line scalar $\xi^3$. Under $Y^3_{\IZ_3}$ they transform as
\begin{equation}
A\to B\to C \to A
\end{equation}
and therefore we must have $A=B=C$, from which the first relation in (\ref{relations-yukawas}) follows. The other relations in (\ref{relations-yukawas}) are proven similarly.

\end{document}